\documentclass[aps, prl, 10pt, twocolumn,footinbib,superscriptaddress, longbibliography]{revtex4-2}

\usepackage[T1]{fontenc}
\usepackage{amsmath}
\usepackage{amssymb}
\usepackage{graphicx}
\usepackage{bm}
\usepackage{lipsum}
\usepackage[dvipsnames]{xcolor}
\usepackage[per-mode=symbol]{siunitx}
\DeclareSIUnit\Molar{\textup{M}}
\DeclareSIUnit\rpm{\textup{rpm}}
\DeclareSIUnit\mpc{\textup{mol}\%}

\usepackage{glossaries}
\setacronymstyle{long-short}
\newacronym{op}{OP}{order parameter}
\newacronym{dopc}{DOPC}{1,2-dioleoyl-\textit{sn}-glycero-3-phosphocholine}
\newacronym{dppc}{DPPC}{1,2-dihexadecanoyl-\textit{sn}-glycero-3-phosphocholine}
\newacronym{dope}{DOPE}{1,2-dioleoyl-\textit{sn}-glycero-3-phosphoethanolamine}
\newacronym{dspe}{DSPE-PEG}{1,2-distearoyl-sn-glycero-3-phosphoethanolamine-N-[amino(polyethylene glycol)-2000]}
\glsunset{dopc}
\glsunset{dppc}
\glsunset{dope}
\glsunset{dspe}
\newacronym{ld}{Ld}{liquid--disordered}
\newacronym{lo}{Lo}{liquid--ordered}
\newacronym{md}{MD}{molecular dynamics}
\newacronym[longplural={giant plasma membrane vesicles}, plural={GPMVs}]{gpmv}{GPMV}{giant plasma membrane vesicle}

\newacronym{dope-biotin}{DOPE--biotin}{1,2-dioleoyl-sn-glycero-3-phosphoethanolamine-N-(cap biotinyl)}
\newacronym{dppe-biotin}{DPPE--biotin}{1,2-dipalmitoyl-sn-glycero-3-phosphoethanolamine-N-(cap biotinyl)}
\newacronym{dspe-peg-biotin}{DSPE--PEG--biotin}{1,2-distearoyl-sn-glycero-3-phosphoethanolamine-N-[biotinyl(polyethyleneglycol)-2000]}

\newcommand{\vecr}{{\bf{r}}}
\newcommand{\vecrho}{{\bm{\rho}}}
\newcommand{\barh}{\bar{h}}
\newcommand{\barphi}{\bar{\phi}}
\newcommand{\kB}{k_\text{B}}
\newcommand{\dd}{\mathrm{d}}
\newcommand{\corrfunc}{\mathfrak{C}}
\newcommand{\BesselJ}{\mathcal{J}}
\newcommand{\iu}{\mathrm{i}}

\begin{document}

\title{Protein induced lipid demixing in homogeneous membranes}
\author{Bernd Henning Stumpf}
\affiliation{PULS Group, Institut f{\"u}r Theoretische Physik, IZNF, Friedrich-Alexander-Universit{\"a}t Erlangen-N{\"u}rnberg, Cauerstra\ss{}e 3, 91058 Erlangen, Germany}

\author{Piotr Nowakowski}
\affiliation{Max-Planck-Institut f{\"u}r Intelligente Systeme Stuttgart, Heisenbergstr. 3, 70569 Stuttgart, Germany}
\affiliation{Institut f{\"u}r Theoretische Physik IV, Universit{\"a}t Stuttgart, Pfaffenwaldring 57, 70569 Stuttgart}

\author{Christian Eggeling}
\affiliation{Institute of Applied Optics and Biophysics, Friedrich-Schiller-University Jena, 07743 Jena, Germany}
\affiliation{Leibniz Institute of Photonic Technology e.V., 07745 Jena, Germany}
\affiliation{MRC Human Immunology Unit and Wolfson Imaging Centre Oxford, MRC Weatherall Institute of Molecular Medicine, University of Oxford, Oxford OX3 9DS, United Kingdom
}
\author{Anna Macio\l{}ek}
\affiliation{Max-Planck-Institut f{\"u}r Intelligente Systeme Stuttgart, Heisenbergstr. 3, 70569 Stuttgart, Germany}
\affiliation{Institute of Physical Chemistry, Polish Academy of Sciences, Kasprzaka 44/52, PL-01-224 Warsaw, Poland}

\author{Ana-Sun\v{c}ana Smith}

\affiliation{PULS Group, Institut f{\"u}r Theoretische Physik, IZNF, Friedrich-Alexander-Universit{\"a}t Erlangen-N{\"u}rnberg, Cauerstra\ss{}e 3, 91058 Erlangen, Germany}
\affiliation{Group of Computational Life Sciences, Department of Physical Chemistry, Ru\dj{}er Bo\v{s}kovi\'c Institute, Bijeni\v{c}ka 54, 10000, Zagreb}
\begin{abstract}
Specific lipid environments are necessary for the establishment of protein signalling platforms in membranes, yet their origin has been highly debated. We present a continuum, exactly solvable model of protein induced local demixing of lipid membranes. The coupling between a local composition and a local thickness of the membrane induces lipid domains around inclusions with hydrophobic mismatch, even for temperatures above the miscibility critical point of the membrane. The model qualitatively explains the experimentally observed formation of lipid domains induced by anchoring of reconstituted actin in flat supported lipid bilayers.
\end{abstract}
\maketitle

The formation of macromolecular platforms involving a number of proteins is the backbone
of many cellular functions. Their assembly is strongly linked to particular lipid
environments~\cite{Ye2016, Nyholm2007}, which requires reorganisation of the heterogeneous
lipid membrane~\cite{Sezgin2017, Helms2004}. The guiding principle for the formation of
macromolecular assemblies is conceptually simple --- different lipid species show varying
interactions with each other or with membrane proteins, resulting in distinct domains of
preferential lipid order and molecular interactions.

It is often considered that cellular membranes are tuned close to the miscibility critical 
point~\cite{Veatch2008,Machta2011}. However, living cellular membranes  show only
nanoscopic domain formation~\cite{Meder2006, Pralle2000}, dynamically forming 
assemblies of tens of nanometers in size~\cite{Levental2020}, as suggested in studies
of reconstituted membranes~\cite{Marx2002}. These observations practically preclude
demixing phase transition as a mechanism of formation of heterogeneous organizations
such as lipid nanodomains (`rafts')
in living cellular membranes. In the mixed phase above the critical temperature
$T_\mathrm{c}$, heterogeneities in the composition can appear due to thermal
fluctuations; thus, they are small in size and do not last long.
Demixing phase transition is found in model membranes, which are lipid mixtures
containing cholesterol~\cite{Marx2002}, as well as  in \glspl{gpmv} where the
lipid and protein makeup is close to cellular levels~\cite{VEATCH20033074, Veatch2008}.

When considering spatial heterogeneities in membranes, one cannot neglect the
role of anchored proteins or lipids.
Perturbations induced by them can have a profound effect on the lipid membrane 
composition and
ordering~\cite{Hammond2005, Stone2017}. In particular, it was suggested that they may
stabilize membrane composition fluctuations above $T_\mathrm{c}$, fostering ordering 
around the perturbation 
centers~\cite{Stone2017}. Furthermore, direct coupling between membrane composition,
spontaneous curvature and protein recruitment was clearly demonstrated
theoretically~\cite{Sens2000, Rautu2015, Ayton2005, Veksler2007, Sadeghi2014} and
experimentally using membrane tethers~\cite{Simunovic11226, Prevost2015}. 
In these systems, proteins with hydrophobic mismatch (different length of the hydrophobic
core as compared to the 
lipids~\cite{Venturoli2005, Domanski2012, Lin2020, *Lin2020c,Schafer2011, Meyer2010}),
attract lipids that fit the spontaneous curvature, thereby building a concave or convex
shape to fill in the height mismatch.

\begin{figure*}[t!]
\includegraphics[width=0.9\textwidth]{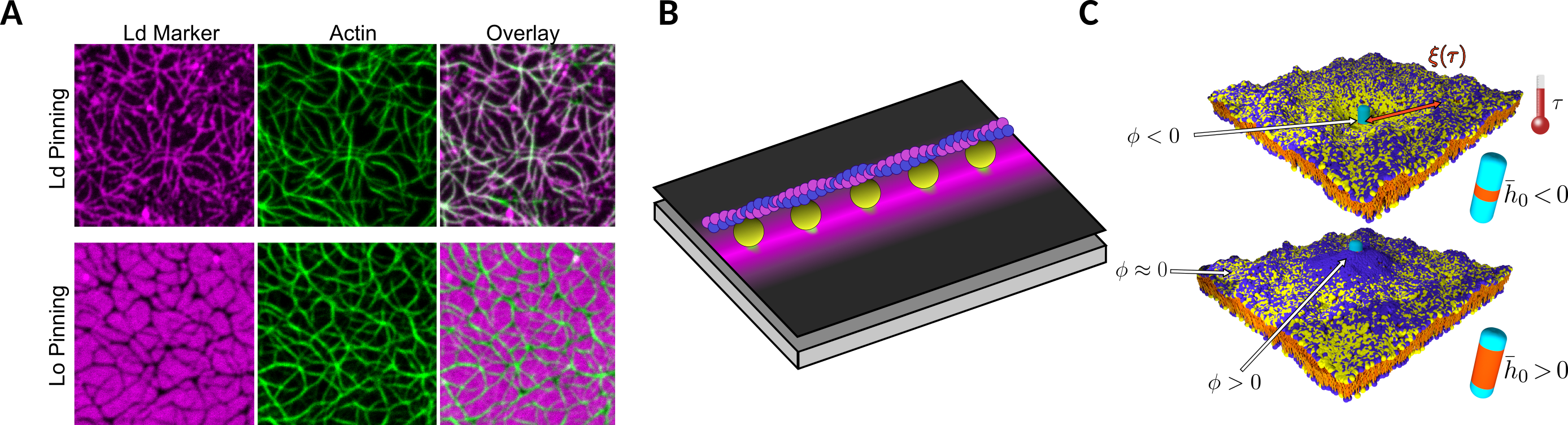}
\caption{\label{fig1} 
    (\textbf{A})~Experimental image of an actin network bounded to \gls{lo} or \gls{ld} anchors (\Gls{lo} or \Gls{ld} Pinning). The membrane was stained by a \gls{ld} marker (magenta) and the actin network by a green marker. The rightmost panel shows an overlay of both markers. The phases induced on the membrane bridge the gaps between the anchors. The image was taken below the critical temperature $T_\mathrm{c}$ of
    a bare membrane (for the control panels see Supplemental Material as well as~\cite{10.7554/eLife.01671}).
    Similar domains have also been observed above $T_\mathrm{c}$~\cite{10.7554/eLife.01671}.
    (\textbf{B})~Schematic illustration of the experimental system. The actin fibre is pinned to 
    the membrane with streptavidins (yellow) that are anchored to either \gls{lo} or 
    \gls{ld} lipids. The anchors induce local demixing (magenta) of the membrane.
    (\textbf{C})~Schematic illustration of the mechanism of formation of the lipid domains in the membrane discussed in this paper. Depending on the sign
    of the excess hydrophobic mismatch $\barh_0$, the lipid composition around the anchor will preferentially be in one of the two phases (\gls{lo} or
    \gls{ld}), represented here by the mixtures of blue and yellow lipids. In the theoretical model, the phase is 
    characterized by the sign of the composition order parameter $\phi$.
    Away from the critical point of demixing, the domain size is of the order of the correlation length $\xi$ at the dimensionless temperature $\tau$ as defined in the text.
}
\end{figure*}

Here, we investigate a complementary mechanism in which mismatched proteins attract lipids
of the right chain length~(Fig.~\ref{fig1}\textbf{C}). So far, this effect was studied only through a
numeric solution of the resulting shape equations~\cite{Shrestha2020}. Here, we present a
fully solvable model that couples the membrane thickness described by Helfrich--like
Hamiltonian~\cite{Bitbol2012} with the composition of the membrane, accounted for by a
functional of a Landau--Ginzburg type. We find that this coupling can recover the formation
of lipid domains around lipids linked to a reconstituted actin cortex
filament~\cite{10.7554/eLife.01671}. 

Specifically, we make a supported lipid bilayer from a mixture of saturated lipids
(\gls{dppc}), unsaturated lipids (\gls{dopc}) and cholesterol in a ratio 
of 35:35:30 mol\% (for experimental details see Supplemental Material). 

Due to the presence of a mica support, the membrane midplane is
essentially flat so
that the spontaneous curvature of the membrane cannot be induced.
Naturally, given that the monolayer profile has a curvature on its
own, additional effects in shaping the boundary between domains
may take place if lipids themselves have curvature preference.
However, these effects are considered to be small, and are hence
neglected in our model.

Below a critical miscibility temperature  $T_\mathrm{c}$, these membranes phase separate
into \gls{lo} and \gls{ld} domains~\cite{VEATCH20033074}. The \gls{lo} phase is
enriched in  cholesterol and saturated lipids, and shows higher extension in the
lipid acyl chains as compared to the \gls{ld} phase enriched in unsaturated 
and less extended lipids~\cite{Brown1998, Bleecker2016}. The perturbation to such a membrane is caused by the reconstituted
actin cytoskeleton, the network--like structure that is locally pinned to the membrane 
through membrane--integrated biotinylated lipids and streptavidin--tagged actin building blocks
(Fig.~\ref{fig1}\textbf{B}). Notably, if streptavidin was coupled to 
saturated or unsaturated lipids, the \gls{lo} or the \gls{ld} phase would 
appear below the actin filaments, respectively (see Fig.~\ref{fig1}\textbf{A}). 
This organisation is observed 
above $T_\mathrm{c}$, as measured in unperturbed membranes (see Supplemental Material, Fig.~1), and persists 
deep below $T_\mathrm{c}$~\cite{10.7554/eLife.01671} (Fig.~\ref{fig1}\textbf{A}). 
The observed domains seem to be a result of local adsorption phenomena rather 
than macroscopic phase separation,
presumably because the immobile lipid anchors destroy the miscibility critical point of the membrane~\cite{Dotsenko1995, Dotsenko2007}.

\begin{figure*}[hbt!]
\includegraphics[width=0.95\textwidth]{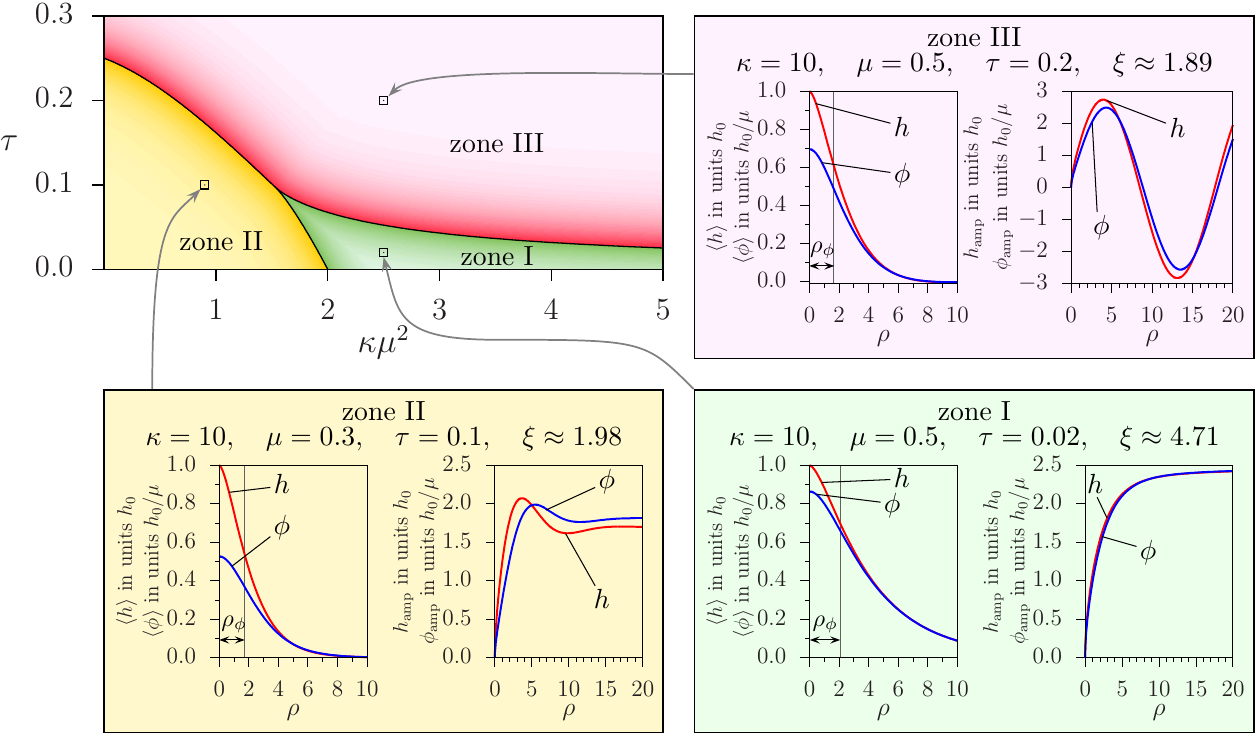}
\caption{\label{fig2} Zones of the model. Top left panel presents the space of parameters. 
The average \glspl{op} $\left<h\right>$ and $\left<\phi\right>$, and their 
amplitudes $h_\text{amp}$ and $\phi_\text{amp}$ multiplying the leading order 
exponential decay for large distances are shown in the other three panels. All 
functions have been calculated for $\kappa=10$, for the points in the space of 
parameters indicated by arrows. Gray vertical lines on the plots of 
average \glspl{op} indicates the size $\rho_\phi$ of the formed domains.}
\end{figure*}
\glslocalreset{op}

To rationalize this formation of domains around immobilised anchored proteins, 
and explain our experimental observations, we devise a minimal theoretical 
model based on two \glspl{op}, where  membrane excess 
thickness $\barh\left(\vecr\right)$ and composition $\barphi\left(\vecr\right)$ are 
coupled~\cite{Grage2016}. We assume that the midlayer of the membrane is flat, and the excess thickness $\barh$ is defined as a difference 
between local and bulk thickness of the bilayer (measured in the direction perpendicular to the midlayer), whereas the composition field $\barphi$
is a difference between local concentration of saturated lipids and their 
concentration at the critical demixing point. With our membrane being 
intrinsically flat, the position $\vecr$ is assumed to be any vector from 
a two--dimensional plane, and the \glspl{op} are allowed to take any real value. 

The energy of the membrane is thus modeled as:
\begin{multline}\label{P:hamiltonianI}
\beta\mathcal{H}_0\left[\barh\left(\vecr\right),\barphi\left(\vecr\right)\right]=\int \mathrm{d}^2 r\Big[ \frac{\sigma}{2} \big( \nabla \barphi\left(\vecr\right)\!\big)^2+t\,\barphi^2\left(\vecr\right)+\\+\frac{\gamma}{2} \big( \barh\left(\vecr\right)-\alpha\, \barphi\left(\vecr\right)\!\big)^2+\frac{\kappa}{2}\big( \nabla^2 \barh\left(\vecr\right)\!\big)^2\Big],
\end{multline}
where $\beta=\left(\kB T\right)^{-1}$. The parameters $\sigma$ and $t$, like in 
the Gaussian model, measure the energy cost of inhomogeneity of the composition 
of the membrane and the reduced temperature, respectively. The model is well 
defined only for $t>0$, and the limit $t\to 0$ corresponds to approaching the 
critical point of demixing  from above. The third term 
in Eq.~\eqref{P:hamiltonianI} describes the coupling between the 
two \glspl{op}. Here, we relate the local composition of the 
lipids $\barphi\left(\vecr\right)$ with the excess thickness of the 
membrane $\barh\left(\vecr\right)$. For simplicity, this relation is 
assumed to be linear with a coefficient $\alpha$. Parameter $\gamma$ regulates 
the energy cost of a deviation from the postulated relation. The final term is 
the elastic energy stored in the deformation of the membrane thickness with 
a bending stiffness $\kappa$. The latter is related to the bending stiffness 
in the Helfrich model as $\kappa = \kappa_\mathcal{H}/4$~\cite{Bitbol2012}. 
In order to reduce the number of free parameters, 
in Hamiltonian \eqref{P:hamiltonianI} we neglect the term $\frac{\eta}{2} \big(\nabla \barh\left(\vecr\right)\!\big)^2$, describing the energy cost of changing of the thickness of the membrane; as we have checked, as long as $\eta/\sqrt{\kappa\gamma}<2$, adding this term is not changing qualitatively the properties of the model \footnote{For example, using estimates for coefficients from \cite{Bories2018}, the value of $\eta/\sqrt{\kappa\gamma}$ does not exceed $0.5$.}.
Moreover, to keep analytic tractability, higher order terms are neglected in both 
the concentration and the thickness fields~\cite{Huang1986}.

The protein anchors are modeled as rigid, point--like 
inclusions~\cite{Netz1997, Dommersnes1999} that have an excess thickness $\barh_0$
\begin{equation}
\beta \mathcal{H}_\text{int}=\frac{\lambda}{2}\sum_{i=1}^N \big(\barh\left(\vecr_i\right)-\barh_0\big)^2,
\end{equation}
where $N$ and $\vecr_i$ denote the number of inclusions and their positions. 
For simplicity we take $\lambda\to\infty$.

Because the Hamiltonian is quadratic in both \glspl{op} and their derivatives, 
it is possible to calculate the partition function, \gls{op} profiles and 
correlation functions analytically using the method of path 
integrals~\cite{Schmidt2012, Janes2019}. To simplify the analysis (see Supplemental Material), we 
introduce the length scale $\zeta=\left(\kappa/\gamma\right)^{1/4}$, associated 
with the bulk membrane in the absence of coupling to the composition OP. Together 
with $\sigma$ it allows us to define the dimensionless 
position $\vecrho=\vecr/\zeta$, 
and \glspl{op}: $h\left(\vecrho\right)=\barh\left(\zeta\vecrho\right)/\zeta$ 
and $\phi\left(\vecrho\right)=\barphi\left(\zeta\vecrho\right) \sigma^{1/2}$. Hence, 
bending stiffness of the membrane $\kappa$, dimensionless reduced 
temperature $\tau=t \zeta^2/\sigma$, and the dimensionless coupling 
between \glspl{op} $\mu=\alpha \sigma^{-1/2}/\zeta$ remain as free parameters. 
Interestingly, $\kappa$ and $\mu$ enter almost all our equations in the 
combination  $\kappa\mu^2$, which we denote by  $\omega$. 

We first discuss the case of a single inclusion ($N=1$), placed 
at $\vecrho_1=\bm{0}$. In this case, due to the rotational symmetry of 
the model, the \gls{op} profiles $\left< h\right>$ and $\left<\phi\right>$ depend 
only on the distance $\rho$ from the origin. To easily extract their long distance 
behavior, it is convenient to decompose the profiles into the amplitude and exponential decay
\begin{subequations}\label{P:decomposition}
\begin{align}
\left<h\right>\left(\rho; \kappa, \mu, \tau\right)&=h_\text{amp}\left(\rho; \kappa, \mu, \tau\right) \exp\left(-\rho/\xi\right)/\sqrt{\rho},\\
\left<\phi\right>\left(\rho; \kappa, \mu, \tau\right)&=\phi_\text{amp}\left(\rho; \kappa, \mu, \tau\right) \exp\left(-\rho/\xi\right)/\sqrt{\rho},
\end{align} 
\end{subequations}
and the amplitudes $h_\text{amp}$ and $\phi_\text{amp}$ are, as a function 
of $\rho$, bounded and not decaying to 0. The parameter $\xi\left(\tau,\omega\right)$ denotes the bulk correlation length, i.e., the lengthscale of decay of the correlation functions. (All three possible two--point correlation functions, $h$--$h$, $h$--$\phi$ and $\phi$--$\phi$, decay on the same lengthscale $\xi$.)
The bulk correlation 
length (in units of $\zeta$)   diverges like $\left(2\tau\right)^{-1/2}$ for $\tau\to 0$ 
and has a limit $\sqrt{2}$ for $\tau\to\infty$.

Depending on the values of parameters $\omega$ and $\tau$, we discover 
three distinct behaviors of the thickness and composition 
profiles (Fig.~\ref{fig2}). For small effective temperatures $\tau$ and 
large $\omega$ (\textit{zone I}) the \gls{op} decay to zero for $\rho\to\infty$ 
with monotonic amplitudes $h_\text{amp}$ and $\phi_\text{amp}$. On the other 
hand, for small $\tau$ and $\omega$ (\textit{zone II}) the amplitudes show some 
decaying oscillations and are typically non--monotonic. The exponential decays 
of \glspl{op}, observed in zones I and II, emerge from the Gaussian model for 
the composition field, dominating at small $\tau$. For 
large $\tau$ (\textit{zone III}) the profiles have a form of damped oscillations, 
which is a typical feature of Helfrich model. Here, the \gls{op} $h$ dominates 
over $\phi$ and induces oscillations for both \glspl{op}. We note that 
crossing the borders between regimes yields a smooth change of 
both \glspl{op}, i.e., no phase transition takes place. The border of zone III with 
other zones is a Fisher--Widom line \cite{Fisher1969, *Fisher2015, Evans1994}.

The characteristic behavior of the \glspl{op} profiles in each of the zones is 
not visible because of the fast exponential decay of $\left<h\right>$ 
and $\left<\phi\right>$, in turn justifying the decomposition 
in Eq.~\eqref{P:decomposition}. In all regimes, notably, the dominant behavior 
is demixing around the anchor at  $\rho=0$, due to adsorption of lipids with a 
matching thickness, which we identify as a formation of a distinct domain. 
We characterise the size $\rho_{\phi}$ of the domain by the inflection point 
of $\left<\phi\right>$ as a function of $\rho$ (Fig.~\ref{fig2}). Away 
from $T_\mathrm{c}$, $\rho_{\phi}$ is of the order of $\xi$, while upon 
approaching $T_\mathrm{c}$ ($\tau \rightarrow 0$), $\rho_{\phi}$ grows fast 
but converges to a finite value, while $\xi$ diverges~\cite{NSMS}. The 
contrast between the composition of the lipid domain and the bulk is defined 
by the intensity of the protein mismatch $\barh_0$, and becomes more pronounced 
upon reducing temperature, as noted in the experiments~\cite{10.7554/eLife.01671}.

\begin{figure}[t!]
\includegraphics[width=0.9\columnwidth]{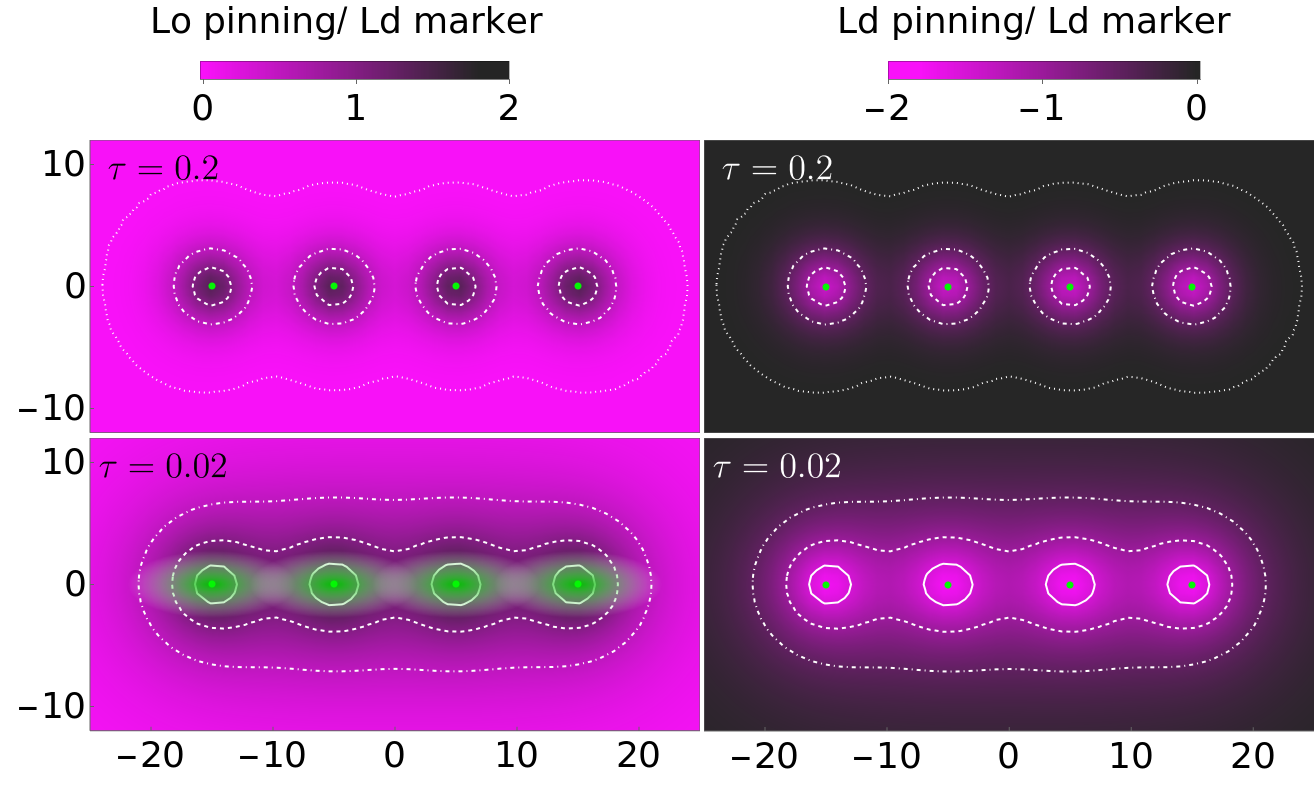}
\caption{\label{fig4}
Mean membrane composition $\left<\phi\right>$ around a
linear arrangement of four membrane
inclusions, calculated from the theoretical model for $\omega=\kappa\mu^2 = 2.5$ and two different values of $\tau$. All lengths are given in units of $\zeta$. 
The positions of the inclusions are marked by green dots.
Contour lines are shown for $\left<\phi\right>=0$ (dotted),
$\left<\phi\right>=\pm 0.5$ (dot--dashed),
$\left<\phi\right>=\pm 1$ (dashed) 
and $\left<\phi\right>=\pm 1.5$ (solid).
The color code is motivated by the colors used in the experiment:
for positive mismatch $\barh_0=+1$ (left column) it represents
the \gls{ld} marker with \gls{lo} pinning, whereas for
negative mismatch $\barh_0=-1$ (right column) it
represents the \gls{ld} marker using \gls{ld} pinning.
The two color codes are used as the homogeneous phase 
in the experiment has a different color depending on the type of pinning.
A possible actin overlay is shown in the lower left panel with green color. Upon approaching the
critical
temperature, the domains of the $\phi$ field around the pinning points 
grow in size, and finally merge forming a bridge 
that connects all the actin pinning points.
}
\end{figure}

To mimic lipid--anchored streptavidin attachments to actin filament as in 
experiments (Fig.~\ref{fig1}), we create an array of $N$ anchors located in 
points $\vecrho_1, \ldots, \vecrho_N$. To find the profiles (see Supplemental Material), 
we first calculate the two correlation functions for the membrane without anchors
\begin{subequations}
\begin{align}
 \corrfunc_{hh}\left(\rho\right)&=\frac{1}{2\pi\kappa}\int_0^\infty\! \frac{x \left(x^2+\omega+2\tau\right) \BesselJ_0\left(\rho\, x\right)}{\left(x^4+1\right)\left(x^2+2\tau+\omega\right)-\omega}\, \dd x,\\
 \corrfunc_{h\phi}\left(\rho\right)&=\frac{\mu}{2\pi}\int_0^\infty \frac{x\, \BesselJ_0\left(\rho\, x\right)}{\left(x^4+1\right)\left(x^2+2\tau+\omega\right)-\omega}\, \dd x,
\end{align}
\end{subequations}
where $\BesselJ_0$ denotes the Bessel function of the first kind of order 0. 
The resulting profiles are given by
\begin{subequations}\label{profilesN}
\begin{align}
 \left<h\right>\left(\vecrho\right)=
 \sum_{k,k'=1}^{N} h_0\, \mathbb{M}^{-1}_{k,k'} \corrfunc_{hh}(\left|\vecrho_{k'}-\vecrho\right|),\\
 \left<\phi\right>\left(\vecrho\right)=
 \sum_{k,k'=1}^{N} h_0\, \mathbb{M}^{-1}_{k,k'} \corrfunc_{h\phi}(\left|\vecrho_{k'}-\vecrho\right|),
\end{align}
\end{subequations}
where the matrix $\mathbb{M}_{k, k'}=\corrfunc_{hh}(\left|\vecrho_k - \vecrho_{k'}\right|)$, 
and $h_0=\barh_0/\zeta$. We note that for $N=1$, Eq.~\eqref{profilesN} reduces 
to Eq.~\eqref{P:decomposition}, and for $\mu=0$ the results equivalent to 
those obtained for a deformed membrane are recovered \cite{Schmidt2012, Bihr2015}.

Fig.~\ref{fig4} shows a representation of the membrane composition 
around a linear array of membrane inclusions, mimicking the actin filament 
pinned by a lipid--protein complex. A positive value of $\left<\phi\right>$ is 
used here to indicate a \gls{lo}--domain, and a negative value 
of $\left<\phi\right>$ indicates a \gls{ld}--domain. Clearly, as the reduced temperature $\tau$ decreases (zones I and II), the amplitude and the range of developing composition and thickness profiles increases, inducing  domains around each inclusion with a  composition different from that in the bulk. The composition of the domain depends on the strength of the hydrophobic mismatch. When the size $\rho_\phi$ of the domains is of the order of the distance between inclusions, the domains coalesce and form one elongated region of increased $\phi$. 
This behavior is in a qualitative agreement with the experimental observations \cite{10.7554/eLife.01671}. Visual inspection of  images for  temperatures higher than $T_{\mathrm{c}}$ (not shown  here) reveals \gls{ld} (\gls{lo})-enriched  domains around streptavidin anchors that are not everywhere completely connected.  Upon reducing the temperature, these domains become more pronounced, grow, and coalesce to copy the structure of the underlying actin network --- see Fig.~\ref{fig1}\textbf{A}.

It is plausible that adding large number of anchors with only positive (negative) hydrophobic mismatch to the membrane in our experiment  completely destroys the critical point of the bare membrane (or, alternatively, the critical point and the line of coexistence between \gls{ld} and \gls{lo} phases is shifted away from the critical concentration of a bare membrane~\cite{Edison2015}, such that they are not visible in the experimental system with anchors). Thus, upon decreasing temperature below $T_{\mathrm{c}}$ of a bare membrane, the system stays in a single phase. The observed structure with domains following the actin network is caused entirely due to adsorption phenomenon induced by the hydrophobic mismatch, where the lipid environment tries to adapt to the height of the streptavidin anchor.

Finally we note that, since the experimental membrane with anchors is in the mixed state, our model is capable of qualitatively explaining the experimental results for the whole range of temperatures. The relation between the reduced temperature $\tau$ in the model and the temperature in the experiment can be established by comparing the correlation lengths.

In conclusion, we have introduced a continuum, exactly solvable model for protein 
induced local demixing of lipid membranes. The novelty in this model lies in the
coupling of membrane composition to the hydrophobic mismatch. Consequently, the 
stress introduced by the mismatched anchor is released by locally accommodating 
composition of the membrane. As such, the anchors serve effectively as an adsorption 
sites for lipids with similar mismatch and the transition to the bulk phase is 
obtained by gradually changing the thickness through the modulating membrane 
composition. This mechanism is complementary to demixing via curvature effects, previously discussed in the literature \cite{Sens2000, Rautu2015, Ayton2005, Veksler2007, Sadeghi2014, Simunovic11226, Prevost2015}. Consequently, both mechanisms should be considered together to fully describe the formation of protein domains in unsupported membranes.

Our Gaussian approach relates to Ising systems, and is thus capable 
of identifying a characteristic length scale $\xi\left(\tau, \omega\right)$ 
and $\zeta$ for the domain formation, which seems to be recovered in experiments. 
Knowing the material constant $\kappa$, the amplitude of the correlation length 
and the distance between pinning sites, we can predict the range of temperatures 
in which, for a given $\mu$,  the domains around separate inclusions are big 
enough to merge into a single one enclosing all pinning points. The 
parameter $\mu$ could be inferred from the amplitude of the \glspl{op} at 
the pinning point. 
Quantitative consistency between experiments and theory, however, is likely to require more detailed 
models for both the composition~\cite{Honerkamp-Smith2008} and the 
thickness fields~\cite{Bitbol2012}. It would be potentially interesting to cast the  effect into the framework of random disorder.

The extension of this model to many dynamic pinning sites can further be used to study
both static and dynamic assemblies of membrane heterogeneities,  where the 
fluctuations will induce protein interactions and affect their organization and 
distributions in the membrane~\cite{ArandaEspiroza1996,Bitbol2012,Bihr2015}. 
Furthermore, the dynamics of these assemblies will be 
determined by the dynamics of the pinning sites, which can bind and unbind from
the membrane, as well as the dynamics of the actin cortex itself. The 
timescales of heterogeneity assembly and disassembly could therefore also serve
as a measure for actin activity and protein binding. These are interesting perspectives 
of this work, that will certainly be explored in the future.

We thank Alf Honigmann for providing us with the experimental image (Fig.~\ref{fig1}\textbf{A}).
C.E.~acknowledges support by the lab of Stefan Hell for taking imaging 
data, and  
funding by the MRC (MC\_UU\_12010/unit 
programmes G0902418 and MC\_UU\_12025) 
and the Deutsche Forschungsgemeinschaft (Jena Excellence 
Cluster ``Balance of the Microverse''; project number 316213987
-- SFB 1278 (project C05)). B.H.S. and A-S.S. thank the German Science Foundation 
project SM 289/8-1, AOBJ: 652939.


\begin{thebibliography}{49}%
\makeatletter
\providecommand \@ifxundefined [1]{%
 \@ifx{#1\undefined}
}%
\providecommand \@ifnum [1]{%
 \ifnum #1\expandafter \@firstoftwo
 \else \expandafter \@secondoftwo
 \fi
}%
\providecommand \@ifx [1]{%
 \ifx #1\expandafter \@firstoftwo
 \else \expandafter \@secondoftwo
 \fi
}%
\providecommand \natexlab [1]{#1}%
\providecommand \enquote  [1]{``#1''}%
\providecommand \bibnamefont  [1]{#1}%
\providecommand \bibfnamefont [1]{#1}%
\providecommand \citenamefont [1]{#1}%
\providecommand \href@noop [0]{\@secondoftwo}%
\providecommand \href [0]{\begingroup \@sanitize@url \@href}%
\providecommand \@href[1]{\@@startlink{#1}\@@href}%
\providecommand \@@href[1]{\endgroup#1\@@endlink}%
\providecommand \@sanitize@url [0]{\catcode `\\12\catcode `\$12\catcode
  `\&12\catcode `\#12\catcode `\^12\catcode `\_12\catcode `\%12\relax}%
\providecommand \@@startlink[1]{}%
\providecommand \@@endlink[0]{}%
\providecommand \url  [0]{\begingroup\@sanitize@url \@url }%
\providecommand \@url [1]{\endgroup\@href {#1}{\urlprefix }}%
\providecommand \urlprefix  [0]{URL }%
\providecommand \Eprint [0]{\href }%
\providecommand \doibase [0]{http://dx.doi.org/}%
\providecommand \selectlanguage [0]{\@gobble}%
\providecommand \bibinfo  [0]{\@secondoftwo}%
\providecommand \bibfield  [0]{\@secondoftwo}%
\providecommand \translation [1]{[#1]}%
\providecommand \BibitemOpen [0]{}%
\providecommand \bibitemStop [0]{}%
\providecommand \bibitemNoStop [0]{.\EOS\space}%
\providecommand \EOS [0]{\spacefactor3000\relax}%
\providecommand \BibitemShut  [1]{\csname bibitem#1\endcsname}%
\let\auto@bib@innerbib\@empty
\bibitem [{\citenamefont {Ye}\ \emph {et~al.}(2016)\citenamefont {Ye},
  \citenamefont {McLean},\ and\ \citenamefont {Sligar}}]{Ye2016}%
  \BibitemOpen
  \bibfield  {author} {\bibinfo {author} {\bibfnamefont {X.}~\bibnamefont
  {Ye}}, \bibinfo {author} {\bibfnamefont {M.~A.}\ \bibnamefont {McLean}}, \
  and\ \bibinfo {author} {\bibfnamefont {S.~G.}\ \bibnamefont {Sligar}},\
  }\bibfield  {title} {\enquote {\bibinfo {title} {Conformational equilibrium
  of talin is regulated by anionic lipids},}\ }\href {\doibase
  10.1016/j.bbamem.2016.05.005} {\bibfield  {journal} {\bibinfo  {journal}
  {Biochim. Biophys. Acta, Biomembr.}\ }\textbf {\bibinfo {volume} {1858}},\
  \bibinfo {pages} {1833--1840} (\bibinfo {year} {2016})}\BibitemShut {NoStop}%
\bibitem [{\citenamefont {Nyholm}\ \emph {et~al.}(2007)\citenamefont {Nyholm},
  \citenamefont {{\"{O}}zdirekcan},\ and\ \citenamefont
  {Killian}}]{Nyholm2007}%
  \BibitemOpen
  \bibfield  {author} {\bibinfo {author} {\bibfnamefont {T.~K.~M.}\
  \bibnamefont {Nyholm}}, \bibinfo {author} {\bibfnamefont {S.}~\bibnamefont
  {{\"{O}}zdirekcan}}, \ and\ \bibinfo {author} {\bibfnamefont {J.~A.}\
  \bibnamefont {Killian}},\ }\bibfield  {title} {\enquote {\bibinfo {title}
  {How protein transmembrane segments sense the lipid environment},}\ }\href
  {\doibase 10.1021/bi061941c} {\bibfield  {journal} {\bibinfo  {journal}
  {Biochem.}\ }\textbf {\bibinfo {volume} {46}},\ \bibinfo {pages} {1457--1465}
  (\bibinfo {year} {2007})}\BibitemShut {NoStop}%
\bibitem [{\citenamefont {Sezgin}\ \emph {et~al.}(2017)\citenamefont {Sezgin},
  \citenamefont {Levental}, \citenamefont {Mayor},\ and\ \citenamefont
  {Eggeling}}]{Sezgin2017}%
  \BibitemOpen
  \bibfield  {author} {\bibinfo {author} {\bibfnamefont {E.}~\bibnamefont
  {Sezgin}}, \bibinfo {author} {\bibfnamefont {I.}~\bibnamefont {Levental}},
  \bibinfo {author} {\bibfnamefont {S.}~\bibnamefont {Mayor}}, \ and\ \bibinfo
  {author} {\bibfnamefont {C.}~\bibnamefont {Eggeling}},\ }\bibfield  {title}
  {\enquote {\bibinfo {title} {{The mystery of membrane organization:
  composition, regulation and roles of lipid rafts}},}\ }\href
  {https://doi.org/10.1038/nrm.2017.16 http://10.0.4.14/nrm.2017.16
  https://www.nature.com/articles/nrm.2017.16#supplementary-information}
  {\bibfield  {journal} {\bibinfo  {journal} {Nat. Rev. Mol. Cell Biol.}\
  }\textbf {\bibinfo {volume} {18}},\ \bibinfo {pages} {361--374} (\bibinfo
  {year} {2017})}\BibitemShut {NoStop}%
\bibitem [{\citenamefont {Helms}\ and\ \citenamefont
  {Zurzolo}(2004)}]{Helms2004}%
  \BibitemOpen
  \bibfield  {author} {\bibinfo {author} {\bibfnamefont {J.~B.}\ \bibnamefont
  {Helms}}\ and\ \bibinfo {author} {\bibfnamefont {C.}~\bibnamefont
  {Zurzolo}},\ }\bibfield  {title} {\enquote {\bibinfo {title} {Lipids as
  targeting signals: Lipid rafts and intracellular trafficking},}\ }\href
  {\doibase 10.1111/j.1600-0854.2004.0181.x} {\bibfield  {journal} {\bibinfo
  {journal} {Traffic}\ }\textbf {\bibinfo {volume} {5}},\ \bibinfo {pages}
  {247--254} (\bibinfo {year} {2004})}\BibitemShut {NoStop}%
\bibitem [{\citenamefont {Veatch}\ \emph {et~al.}(2008)\citenamefont {Veatch},
  \citenamefont {Cicuta}, \citenamefont {Sengupta}, \citenamefont
  {Honerkamp-Smith}, \citenamefont {Holowka},\ and\ \citenamefont
  {Baird}}]{Veatch2008}%
  \BibitemOpen
  \bibfield  {author} {\bibinfo {author} {\bibfnamefont {S.~L.}\ \bibnamefont
  {Veatch}}, \bibinfo {author} {\bibfnamefont {P.}~\bibnamefont {Cicuta}},
  \bibinfo {author} {\bibfnamefont {P.}~\bibnamefont {Sengupta}}, \bibinfo
  {author} {\bibfnamefont {A.}~\bibnamefont {Honerkamp-Smith}}, \bibinfo
  {author} {\bibfnamefont {D.}~\bibnamefont {Holowka}}, \ and\ \bibinfo
  {author} {\bibfnamefont {B.}~\bibnamefont {Baird}},\ }\bibfield  {title}
  {\enquote {\bibinfo {title} {Critical fluctuations in plasma membrane
  vesicles},}\ }\href {\doibase 10.1021/cb800012x} {\bibfield  {journal}
  {\bibinfo  {journal} {ACS Chem. Biol.}\ }\textbf {\bibinfo {volume} {3}},\
  \bibinfo {pages} {287--293} (\bibinfo {year} {2008})}\BibitemShut {NoStop}%
\bibitem [{\citenamefont {Machta}\ \emph {et~al.}(2011)\citenamefont {Machta},
  \citenamefont {Papanikolaou}, \citenamefont {Sethna},\ and\ \citenamefont
  {Veatch}}]{Machta2011}%
  \BibitemOpen
  \bibfield  {author} {\bibinfo {author} {\bibfnamefont {B.~B.}\ \bibnamefont
  {Machta}}, \bibinfo {author} {\bibfnamefont {S.}~\bibnamefont
  {Papanikolaou}}, \bibinfo {author} {\bibfnamefont {J.~P.}\ \bibnamefont
  {Sethna}}, \ and\ \bibinfo {author} {\bibfnamefont {S.~L.}\ \bibnamefont
  {Veatch}},\ }\bibfield  {title} {\enquote {\bibinfo {title} {Minimal model of
  plasma membrane heterogeneity requires coupling cortical actin to
  criticality},}\ }\href {\doibase 10.1016/j.bpj.2011.02.029} {\bibfield
  {journal} {\bibinfo  {journal} {Biophys. J.}\ }\textbf {\bibinfo {volume}
  {100}},\ \bibinfo {pages} {1668--1677} (\bibinfo {year} {2011})}\BibitemShut
  {NoStop}%
\bibitem [{\citenamefont {Meder}\ \emph {et~al.}(2006)\citenamefont {Meder},
  \citenamefont {Moreno}, \citenamefont {Verkade}, \citenamefont {Vaz},\ and\
  \citenamefont {Simons}}]{Meder2006}%
  \BibitemOpen
  \bibfield  {author} {\bibinfo {author} {\bibfnamefont {D.}~\bibnamefont
  {Meder}}, \bibinfo {author} {\bibfnamefont {M.~J.}\ \bibnamefont {Moreno}},
  \bibinfo {author} {\bibfnamefont {P.}~\bibnamefont {Verkade}}, \bibinfo
  {author} {\bibfnamefont {W.~L.~C.}\ \bibnamefont {Vaz}}, \ and\ \bibinfo
  {author} {\bibfnamefont {K.}~\bibnamefont {Simons}},\ }\bibfield  {title}
  {\enquote {\bibinfo {title} {Phase coexistence and connectivity in the apical
  membrane of polarized epithelial cells},}\ }\href {\doibase
  10.1073/pnas.0509885103} {\bibfield  {journal} {\bibinfo  {journal} {Proc.
  Natl. Acad. Sci. U.S.A.}\ }\textbf {\bibinfo {volume} {103}},\ \bibinfo
  {pages} {329--334} (\bibinfo {year} {2006})}\BibitemShut {NoStop}%
\bibitem [{\citenamefont {Pralle}\ \emph {et~al.}(2000)\citenamefont {Pralle},
  \citenamefont {Keller}, \citenamefont {Florin}, \citenamefont {Simons},\ and\
  \citenamefont {H{\"{o}}rber}}]{Pralle2000}%
  \BibitemOpen
  \bibfield  {author} {\bibinfo {author} {\bibfnamefont {A.}~\bibnamefont
  {Pralle}}, \bibinfo {author} {\bibfnamefont {P.}~\bibnamefont {Keller}},
  \bibinfo {author} {\bibfnamefont {E.-L.}\ \bibnamefont {Florin}}, \bibinfo
  {author} {\bibfnamefont {K.}~\bibnamefont {Simons}}, \ and\ \bibinfo {author}
  {\bibfnamefont {J.}~\bibnamefont {H{\"{o}}rber}},\ }\bibfield  {title}
  {\enquote {\bibinfo {title} {Sphingolipid–cholesterol rafts diffuse as
  small entities in the plasma membrane of mammalian cells},}\ }\href {\doibase
  10.1083/jcb.148.5.997} {\bibfield  {journal} {\bibinfo  {journal} {J. Cell
  Biol.}\ }\textbf {\bibinfo {volume} {148}},\ \bibinfo {pages} {997--1008}
  (\bibinfo {year} {2000})}\BibitemShut {NoStop}%
\bibitem [{\citenamefont {Levental}\ \emph {et~al.}(2020)\citenamefont
  {Levental}, \citenamefont {Levental},\ and\ \citenamefont
  {Heberle}}]{Levental2020}%
  \BibitemOpen
  \bibfield  {author} {\bibinfo {author} {\bibfnamefont {I.}~\bibnamefont
  {Levental}}, \bibinfo {author} {\bibfnamefont {K.~R.}\ \bibnamefont
  {Levental}}, \ and\ \bibinfo {author} {\bibfnamefont {F.~A.}\ \bibnamefont
  {Heberle}},\ }\bibfield  {title} {\enquote {\bibinfo {title} {Lipid rafts:
  Controversies resolved, mysteries remain},}\ }\href {\doibase
  10.1016/j.tcb.2020.01.009} {\bibfield  {journal} {\bibinfo  {journal} {Trends
  Cell Biol.}\ }\textbf {\bibinfo {volume} {30}},\ \bibinfo {pages} {341--353}
  (\bibinfo {year} {2020})}\BibitemShut {NoStop}%
\bibitem [{\citenamefont {Marx}\ \emph {et~al.}(2002)\citenamefont {Marx},
  \citenamefont {Schilling}, \citenamefont {Sackmann},\ and\ \citenamefont
  {Bruinsma}}]{Marx2002}%
  \BibitemOpen
  \bibfield  {author} {\bibinfo {author} {\bibfnamefont {S.}~\bibnamefont
  {Marx}}, \bibinfo {author} {\bibfnamefont {J.}~\bibnamefont {Schilling}},
  \bibinfo {author} {\bibfnamefont {E.}~\bibnamefont {Sackmann}}, \ and\
  \bibinfo {author} {\bibfnamefont {R.}~\bibnamefont {Bruinsma}},\ }\bibfield
  {title} {\enquote {\bibinfo {title} {Helfrich repulsion and dynamical phase
  separation of multicomponent lipid bilayers},}\ }\href {\doibase
  10.1103/PhysRevLett.88.138102} {\bibfield  {journal} {\bibinfo  {journal}
  {Phys. Rev. Lett.}\ }\textbf {\bibinfo {volume} {88}},\ \bibinfo {pages}
  {138102} (\bibinfo {year} {2002})}\BibitemShut {NoStop}%
\bibitem [{\citenamefont {Veatch}\ and\ \citenamefont
  {Keller}(2003)}]{VEATCH20033074}%
  \BibitemOpen
  \bibfield  {author} {\bibinfo {author} {\bibfnamefont {S.~L.}\ \bibnamefont
  {Veatch}}\ and\ \bibinfo {author} {\bibfnamefont {S.~L.}\ \bibnamefont
  {Keller}},\ }\bibfield  {title} {\enquote {\bibinfo {title} {Separation of
  liquid phases in giant vesicles of ternary mixtures of phospholipids and
  cholesterol},}\ }\href {\doibase
  https://doi.org/10.1016/S0006-3495(03)74726-2} {\bibfield  {journal}
  {\bibinfo  {journal} {Biophys. J.}\ }\textbf {\bibinfo {volume} {85}},\
  \bibinfo {pages} {3074--3083} (\bibinfo {year} {2003})}\BibitemShut {NoStop}%
\bibitem [{\citenamefont {Hammond}\ \emph {et~al.}(2005)\citenamefont
  {Hammond}, \citenamefont {Heberle}, \citenamefont {Baumgart}, \citenamefont
  {Holowka}, \citenamefont {Baird},\ and\ \citenamefont
  {Feigenson}}]{Hammond2005}%
  \BibitemOpen
  \bibfield  {author} {\bibinfo {author} {\bibfnamefont {A.~T.}\ \bibnamefont
  {Hammond}}, \bibinfo {author} {\bibfnamefont {F.~A.}\ \bibnamefont
  {Heberle}}, \bibinfo {author} {\bibfnamefont {T.}~\bibnamefont {Baumgart}},
  \bibinfo {author} {\bibfnamefont {D.}~\bibnamefont {Holowka}}, \bibinfo
  {author} {\bibfnamefont {B.}~\bibnamefont {Baird}}, \ and\ \bibinfo {author}
  {\bibfnamefont {G.~W.}\ \bibnamefont {Feigenson}},\ }\bibfield  {title}
  {\enquote {\bibinfo {title} {Crosslinking a lipid raft component triggers
  liquid ordered-liquid disordered phase separation in model plasma
  membranes},}\ }\href {\doibase 10.1073/pnas.0405654102} {\bibfield  {journal}
  {\bibinfo  {journal} {Proc. Natl. Acad. Sci. U.S.A.}\ }\textbf {\bibinfo
  {volume} {102}},\ \bibinfo {pages} {6320--6325} (\bibinfo {year}
  {2005})}\BibitemShut {NoStop}%
\bibitem [{\citenamefont {Stone}\ \emph {et~al.}(2017)\citenamefont {Stone},
  \citenamefont {Shelby}, \citenamefont {N{\'{u}}{\~{n}}ez}, \citenamefont
  {Wisser},\ and\ \citenamefont {Veatch}}]{Stone2017}%
  \BibitemOpen
  \bibfield  {author} {\bibinfo {author} {\bibfnamefont {M.~B.}\ \bibnamefont
  {Stone}}, \bibinfo {author} {\bibfnamefont {S.~A.}\ \bibnamefont {Shelby}},
  \bibinfo {author} {\bibfnamefont {M.~F.}\ \bibnamefont {N{\'{u}}{\~{n}}ez}},
  \bibinfo {author} {\bibfnamefont {K.}~\bibnamefont {Wisser}}, \ and\ \bibinfo
  {author} {\bibfnamefont {S.~L.}\ \bibnamefont {Veatch}},\ }\bibfield  {title}
  {\enquote {\bibinfo {title} {Protein sorting by lipid phase-like domains
  supports emergent signaling function in {B} lymphocyte plasma membranes},}\
  }\href {\doibase 10.7554/eLife.19891} {\bibfield  {journal} {\bibinfo
  {journal} {eLife}\ }\textbf {\bibinfo {volume} {6}},\ \bibinfo {pages}
  {e19891} (\bibinfo {year} {2017})}\BibitemShut {NoStop}%
\bibitem [{\citenamefont {Sens}\ and\ \citenamefont {Safran}(2000)}]{Sens2000}%
  \BibitemOpen
  \bibfield  {author} {\bibinfo {author} {\bibfnamefont {P.}~\bibnamefont
  {Sens}}\ and\ \bibinfo {author} {\bibfnamefont {S.~A.}\ \bibnamefont
  {Safran}},\ }\bibfield  {title} {\enquote {\bibinfo {title} {Inclusions
  induced phase separation in mixed lipid film},}\ }\href {\doibase
  10.1007/s101890050026} {\bibfield  {journal} {\bibinfo  {journal} {Eur. Phys.
  J. E}\ }\textbf {\bibinfo {volume} {1}},\ \bibinfo {pages} {237--248}
  (\bibinfo {year} {2000})}\BibitemShut {NoStop}%
\bibitem [{\citenamefont {Rautu}\ \emph {et~al.}(2015)\citenamefont {Rautu},
  \citenamefont {Rowlands},\ and\ \citenamefont {Turner}}]{Rautu2015}%
  \BibitemOpen
  \bibfield  {author} {\bibinfo {author} {\bibfnamefont {S.~A.}\ \bibnamefont
  {Rautu}}, \bibinfo {author} {\bibfnamefont {G.}~\bibnamefont {Rowlands}}, \
  and\ \bibinfo {author} {\bibfnamefont {M.~S.}\ \bibnamefont {Turner}},\
  }\bibfield  {title} {\enquote {\bibinfo {title} {Membrane composition
  variation and underdamped mechanics near transmembrane proteins and coats},}\
  }\href {\doibase 10.1103/physrevlett.114.098101} {\bibfield  {journal}
  {\bibinfo  {journal} {Phys. Rev. Lett.}\ }\textbf {\bibinfo {volume} {114}},\
  \bibinfo {pages} {098101} (\bibinfo {year} {2015})}\BibitemShut {NoStop}%
\bibitem [{\citenamefont {Ayton}\ \emph {et~al.}(2005)\citenamefont {Ayton},
  \citenamefont {McWhirter}, \citenamefont {McMurtry},\ and\ \citenamefont
  {Voth}}]{Ayton2005}%
  \BibitemOpen
  \bibfield  {author} {\bibinfo {author} {\bibfnamefont {G.~S.}\ \bibnamefont
  {Ayton}}, \bibinfo {author} {\bibfnamefont {J.~L.}\ \bibnamefont
  {McWhirter}}, \bibinfo {author} {\bibfnamefont {P.}~\bibnamefont {McMurtry}},
  \ and\ \bibinfo {author} {\bibfnamefont {G.~A.}\ \bibnamefont {Voth}},\
  }\bibfield  {title} {\enquote {\bibinfo {title} {Coupling field theory with
  continuum mechanics: A simulation of domain formation in giant unilamellar
  vesicles},}\ }\href {\doibase 10.1529/biophysj.105.059436} {\bibfield
  {journal} {\bibinfo  {journal} {Biophys. J.}\ }\textbf {\bibinfo {volume}
  {88}},\ \bibinfo {pages} {3855--3869} (\bibinfo {year} {2005})}\BibitemShut
  {NoStop}%
\bibitem [{\citenamefont {Veksler}\ and\ \citenamefont
  {Gov}(2007)}]{Veksler2007}%
  \BibitemOpen
  \bibfield  {author} {\bibinfo {author} {\bibfnamefont {A.}~\bibnamefont
  {Veksler}}\ and\ \bibinfo {author} {\bibfnamefont {N.~S.}\ \bibnamefont
  {Gov}},\ }\bibfield  {title} {\enquote {\bibinfo {title} {Phase transitions
  of the coupled membrane-cytoskeleton modify cellular shape},}\ }\href
  {\doibase 10.1529/biophysj.107.113282} {\bibfield  {journal} {\bibinfo
  {journal} {Biophys. J.}\ }\textbf {\bibinfo {volume} {93}},\ \bibinfo {pages}
  {3798--3810} (\bibinfo {year} {2007})}\BibitemShut {NoStop}%
\bibitem [{\citenamefont {Sadeghi}\ \emph {et~al.}(2014)\citenamefont
  {Sadeghi}, \citenamefont {M{\"{u}}ller},\ and\ \citenamefont
  {Vink}}]{Sadeghi2014}%
  \BibitemOpen
  \bibfield  {author} {\bibinfo {author} {\bibfnamefont {S.}~\bibnamefont
  {Sadeghi}}, \bibinfo {author} {\bibfnamefont {M.}~\bibnamefont
  {M{\"{u}}ller}}, \ and\ \bibinfo {author} {\bibfnamefont {R.~L.~C.}\
  \bibnamefont {Vink}},\ }\bibfield  {title} {\enquote {\bibinfo {title} {Raft
  formation in lipid bilayers coupled to curvature},}\ }\href {\doibase
  10.1016/j.bpj.2014.07.072} {\bibfield  {journal} {\bibinfo  {journal}
  {Biophys. J.}\ }\textbf {\bibinfo {volume} {107}},\ \bibinfo {pages}
  {1591--1600} (\bibinfo {year} {2014})}\BibitemShut {NoStop}%
\bibitem [{\citenamefont {Simunovic}\ \emph {et~al.}(2016)\citenamefont
  {Simunovic}, \citenamefont {Evergren}, \citenamefont {Golushko},
  \citenamefont {Pr{\'e}vost}, \citenamefont {Renard}, \citenamefont
  {Johannes}, \citenamefont {McMahon}, \citenamefont {Lorman}, \citenamefont
  {Voth},\ and\ \citenamefont {Bassereau}}]{Simunovic11226}%
  \BibitemOpen
  \bibfield  {author} {\bibinfo {author} {\bibfnamefont {M.}~\bibnamefont
  {Simunovic}}, \bibinfo {author} {\bibfnamefont {E.}~\bibnamefont {Evergren}},
  \bibinfo {author} {\bibfnamefont {I.}~\bibnamefont {Golushko}}, \bibinfo
  {author} {\bibfnamefont {C.}~\bibnamefont {Pr{\'e}vost}}, \bibinfo {author}
  {\bibfnamefont {H.-F.}\ \bibnamefont {Renard}}, \bibinfo {author}
  {\bibfnamefont {L.}~\bibnamefont {Johannes}}, \bibinfo {author}
  {\bibfnamefont {H.~T.}\ \bibnamefont {McMahon}}, \bibinfo {author}
  {\bibfnamefont {V.}~\bibnamefont {Lorman}}, \bibinfo {author} {\bibfnamefont
  {G.~A.}\ \bibnamefont {Voth}}, \ and\ \bibinfo {author} {\bibfnamefont
  {P.}~\bibnamefont {Bassereau}},\ }\bibfield  {title} {\enquote {\bibinfo
  {title} {How curvature-generating proteins build scaffolds on membrane
  nanotubes},}\ }\href {\doibase 10.1073/pnas.1606943113} {\bibfield  {journal}
  {\bibinfo  {journal} {Proc. Natl. Acad. Sci. USA}\ }\textbf {\bibinfo
  {volume} {113}},\ \bibinfo {pages} {11226--11231} (\bibinfo {year}
  {2016})}\BibitemShut {NoStop}%
\bibitem [{\citenamefont {Pr\'evost}\ \emph {et~al.}(2015)\citenamefont
  {Pr\'evost}, \citenamefont {Zhao}, \citenamefont {Manzi}, \citenamefont
  {Lemichez}, \citenamefont {Lappalainen}, \citenamefont {Callan-Jones},\ and\
  \citenamefont {Bassereau}}]{Prevost2015}%
  \BibitemOpen
  \bibfield  {author} {\bibinfo {author} {\bibfnamefont {C.}~\bibnamefont
  {Pr\'evost}}, \bibinfo {author} {\bibfnamefont {H.}~\bibnamefont {Zhao}},
  \bibinfo {author} {\bibfnamefont {J.}~\bibnamefont {Manzi}}, \bibinfo
  {author} {\bibfnamefont {E.}~\bibnamefont {Lemichez}}, \bibinfo {author}
  {\bibfnamefont {P.}~\bibnamefont {Lappalainen}}, \bibinfo {author}
  {\bibfnamefont {A.}~\bibnamefont {Callan-Jones}}, \ and\ \bibinfo {author}
  {\bibfnamefont {P.}~\bibnamefont {Bassereau}},\ }\bibfield  {title} {\enquote
  {\bibinfo {title} {{IRSp53} senses negative membrane curvature and phase
  separates along membrane tubules},}\ }\href {\doibase 10.1038/ncomms9529}
  {\bibfield  {journal} {\bibinfo  {journal} {Nat. Commun.}\ }\textbf {\bibinfo
  {volume} {6}},\ \bibinfo {pages} {8529} (\bibinfo {year} {2015})}\BibitemShut
  {NoStop}%
\bibitem [{\citenamefont {Venturoli}\ \emph {et~al.}(2005)\citenamefont
  {Venturoli}, \citenamefont {Smit},\ and\ \citenamefont
  {Sperotto}}]{Venturoli2005}%
  \BibitemOpen
  \bibfield  {author} {\bibinfo {author} {\bibfnamefont {M.}~\bibnamefont
  {Venturoli}}, \bibinfo {author} {\bibfnamefont {B.}~\bibnamefont {Smit}}, \
  and\ \bibinfo {author} {\bibfnamefont {M.~M.}\ \bibnamefont {Sperotto}},\
  }\bibfield  {title} {\enquote {\bibinfo {title} {Simulation studies of
  protein-induced bilayer deformations, and lipid-induced protein tilting, on a
  mesoscopic model for lipid bilayers with embedded proteins},}\ }\href
  {\doibase 10.1529/biophysj.104.050849} {\bibfield  {journal} {\bibinfo
  {journal} {Biophys. J.}\ }\textbf {\bibinfo {volume} {88}},\ \bibinfo {pages}
  {1778--1798} (\bibinfo {year} {2005})}\BibitemShut {NoStop}%
\bibitem [{\citenamefont {Doma{\'{n}}ski}\ \emph {et~al.}(2012)\citenamefont
  {Doma{\'{n}}ski}, \citenamefont {Marrink},\ and\ \citenamefont
  {Sch{\"{a}}fer}}]{Domanski2012}%
  \BibitemOpen
  \bibfield  {author} {\bibinfo {author} {\bibfnamefont {J.}~\bibnamefont
  {Doma{\'{n}}ski}}, \bibinfo {author} {\bibfnamefont {S.~J.}\ \bibnamefont
  {Marrink}}, \ and\ \bibinfo {author} {\bibfnamefont {L.~V.}\ \bibnamefont
  {Sch{\"{a}}fer}},\ }\bibfield  {title} {\enquote {\bibinfo {title}
  {Transmembrane helices can induce domain formation in crowded model
  membranes},}\ }\href {\doibase 10.1016/j.bbamem.2011.08.021} {\bibfield
  {journal} {\bibinfo  {journal} {Biochim. Biophys. Acta, Biomembr.}\ }\textbf
  {\bibinfo {volume} {1818}},\ \bibinfo {pages} {984--994} (\bibinfo {year}
  {2012})}\BibitemShut {NoStop}%
\bibitem [{\citenamefont {Lin}\ \emph {et~al.}(2020{\natexlab{a}})\citenamefont
  {Lin}, \citenamefont {Lin},\ and\ \citenamefont {Gu}}]{Lin2020}%
  \BibitemOpen
  \bibfield  {author} {\bibinfo {author} {\bibfnamefont {X.}~\bibnamefont
  {Lin}}, \bibinfo {author} {\bibfnamefont {X.}~\bibnamefont {Lin}}, \ and\
  \bibinfo {author} {\bibfnamefont {N.}~\bibnamefont {Gu}},\ }\bibfield
  {title} {\enquote {\bibinfo {title} {Optimization of hydrophobic
  nanoparticles to better target lipid rafts with molecular dynamics
  simulations},}\ }\href {\doibase 10.1039/c9nr09226a} {\bibfield  {journal}
  {\bibinfo  {journal} {Nanoscale}\ }\textbf {\bibinfo {volume} {12}},\
  \bibinfo {pages} {4101--4109} (\bibinfo {year}
  {2020}{\natexlab{a}})}\BibitemShut {NoStop}%
\bibitem [{\citenamefont {Lin}\ \emph {et~al.}(2020{\natexlab{b}})\citenamefont
  {Lin}, \citenamefont {Lin},\ and\ \citenamefont {Gu}}]{Lin2020c}%
  \BibitemOpen
  \bibfield  {author} {\bibinfo {author} {\bibfnamefont {X.}~\bibnamefont
  {Lin}}, \bibinfo {author} {\bibfnamefont {X.}~\bibnamefont {Lin}}, \ and\
  \bibinfo {author} {\bibfnamefont {N.}~\bibnamefont {Gu}},\ }\bibfield
  {title} {\enquote {\bibinfo {title} {Correction: Optimization of hydrophobic
  nanoparticles to better target lipid rafts with molecular dynamics
  simulations},}\ }\href {\doibase 10.1039/D0NR90171G} {\bibfield  {journal}
  {\bibinfo  {journal} {Nanoscale}\ }\textbf {\bibinfo {volume} {12}},\
  \bibinfo {pages} {16389--16389} (\bibinfo {year}
  {2020}{\natexlab{b}})}\BibitemShut {NoStop}%
\bibitem [{\citenamefont {Sch{\"{a}}fer}\ \emph {et~al.}(2011)\citenamefont
  {Sch{\"{a}}fer}, \citenamefont {de~Jong}, \citenamefont {Holt}, \citenamefont
  {Rzepiela}, \citenamefont {de~Vries}, \citenamefont {Poolman}, \citenamefont
  {Killian},\ and\ \citenamefont {Marrink}}]{Schafer2011}%
  \BibitemOpen
  \bibfield  {author} {\bibinfo {author} {\bibfnamefont {L.~V.}\ \bibnamefont
  {Sch{\"{a}}fer}}, \bibinfo {author} {\bibfnamefont {D.~H.}\ \bibnamefont
  {de~Jong}}, \bibinfo {author} {\bibfnamefont {A.}~\bibnamefont {Holt}},
  \bibinfo {author} {\bibfnamefont {A.~J.}\ \bibnamefont {Rzepiela}}, \bibinfo
  {author} {\bibfnamefont {A.~H.}\ \bibnamefont {de~Vries}}, \bibinfo {author}
  {\bibfnamefont {B.}~\bibnamefont {Poolman}}, \bibinfo {author} {\bibfnamefont
  {J.~A.}\ \bibnamefont {Killian}}, \ and\ \bibinfo {author} {\bibfnamefont
  {S.~J.}\ \bibnamefont {Marrink}},\ }\bibfield  {title} {\enquote {\bibinfo
  {title} {Lipid packing drives the segregation of transmembrane helices into
  disordered lipid domains in model membranes},}\ }\href {\doibase
  10.1073/pnas.1009362108} {\bibfield  {journal} {\bibinfo  {journal} {Proc.
  Natl. Acad. Sci. U.S.A.}\ }\textbf {\bibinfo {volume} {108}},\ \bibinfo
  {pages} {1343--1348} (\bibinfo {year} {2011})}\BibitemShut {NoStop}%
\bibitem [{\citenamefont {de~Meyer}\ \emph {et~al.}(2010)\citenamefont
  {de~Meyer}, \citenamefont {Rodgers}, \citenamefont {Willems},\ and\
  \citenamefont {Smit}}]{Meyer2010}%
  \BibitemOpen
  \bibfield  {author} {\bibinfo {author} {\bibfnamefont {F.~J.-M.}\
  \bibnamefont {de~Meyer}}, \bibinfo {author} {\bibfnamefont {J.~M.}\
  \bibnamefont {Rodgers}}, \bibinfo {author} {\bibfnamefont {T.~F.}\
  \bibnamefont {Willems}}, \ and\ \bibinfo {author} {\bibfnamefont
  {B.}~\bibnamefont {Smit}},\ }\bibfield  {title} {\enquote {\bibinfo {title}
  {Molecular simulation of the effect of cholesterol on lipid-mediated
  protein-protein interactions},}\ }\href {\doibase 10.1016/j.bpj.2010.09.030}
  {\bibfield  {journal} {\bibinfo  {journal} {Biophys. J.}\ }\textbf {\bibinfo
  {volume} {99}},\ \bibinfo {pages} {3629--3638} (\bibinfo {year}
  {2010})}\BibitemShut {NoStop}%
\bibitem [{\citenamefont {Shrestha}\ \emph {et~al.}(2020)\citenamefont
  {Shrestha}, \citenamefont {Kahraman},\ and\ \citenamefont
  {Haselwandter}}]{Shrestha2020}%
  \BibitemOpen
  \bibfield  {author} {\bibinfo {author} {\bibfnamefont {A.}~\bibnamefont
  {Shrestha}}, \bibinfo {author} {\bibfnamefont {O.}~\bibnamefont {Kahraman}},
  \ and\ \bibinfo {author} {\bibfnamefont {C.~A.}\ \bibnamefont
  {Haselwandter}},\ }\bibfield  {title} {\enquote {\bibinfo {title} {Regulation
  of membrane proteins through local heterogeneity in lipid bilayer
  thickness},}\ }\href {\doibase 10.1103/physreve.102.060401} {\bibfield
  {journal} {\bibinfo  {journal} {Phys. Rev. E}\ }\textbf {\bibinfo {volume}
  {102}},\ \bibinfo {pages} {060401(R)} (\bibinfo {year} {2020})}\BibitemShut
  {NoStop}%
\bibitem [{\citenamefont {Bitbol}\ \emph {et~al.}(2012)\citenamefont {Bitbol},
  \citenamefont {Constantin},\ and\ \citenamefont {Fournier}}]{Bitbol2012}%
  \BibitemOpen
  \bibfield  {author} {\bibinfo {author} {\bibfnamefont {A.-F.}\ \bibnamefont
  {Bitbol}}, \bibinfo {author} {\bibfnamefont {D.}~\bibnamefont {Constantin}},
  \ and\ \bibinfo {author} {\bibfnamefont {J.-B.}\ \bibnamefont {Fournier}},\
  }\bibfield  {title} {\enquote {\bibinfo {title} {Bilayer elasticity at the
  nanoscale: The need for new terms},}\ }\href {\doibase
  10.1371/journal.pone.0048306} {\bibfield  {journal} {\bibinfo  {journal}
  {PLoS ONE}\ }\textbf {\bibinfo {volume} {7}},\ \bibinfo {pages} {e48306}
  (\bibinfo {year} {2012})}\BibitemShut {NoStop}%
\bibitem [{\citenamefont {Honigmann}\ \emph {et~al.}(2014)\citenamefont
  {Honigmann}, \citenamefont {Sadeghi}, \citenamefont {Keller}, \citenamefont
  {Hell}, \citenamefont {Eggeling},\ and\ \citenamefont
  {Vink}}]{10.7554/eLife.01671}%
  \BibitemOpen
  \bibfield  {author} {\bibinfo {author} {\bibfnamefont {A.}~\bibnamefont
  {Honigmann}}, \bibinfo {author} {\bibfnamefont {S.}~\bibnamefont {Sadeghi}},
  \bibinfo {author} {\bibfnamefont {J.}~\bibnamefont {Keller}}, \bibinfo
  {author} {\bibfnamefont {S.~W.}\ \bibnamefont {Hell}}, \bibinfo {author}
  {\bibfnamefont {C.}~\bibnamefont {Eggeling}}, \ and\ \bibinfo {author}
  {\bibfnamefont {R.}~\bibnamefont {Vink}},\ }\bibfield  {title} {\enquote
  {\bibinfo {title} {A lipid bound actin meshwork organizes liquid phase
  separation in model membranes},}\ }\href {\doibase 10.7554/eLife.01671}
  {\bibfield  {journal} {\bibinfo  {journal} {eLife}\ }\textbf {\bibinfo
  {volume} {3}},\ \bibinfo {pages} {e01671} (\bibinfo {year}
  {2014})}\BibitemShut {NoStop}%
\bibitem [{\citenamefont {Brown}\ and\ \citenamefont
  {London}(1998)}]{Brown1998}%
  \BibitemOpen
  \bibfield  {author} {\bibinfo {author} {\bibfnamefont {D.~A.}\ \bibnamefont
  {Brown}}\ and\ \bibinfo {author} {\bibfnamefont {E.}~\bibnamefont {London}},\
  }\bibfield  {title} {\enquote {\bibinfo {title} {Structure and origin of
  ordered lipid domains in biological membranes},}\ }\href {\doibase
  10.1007/s002329900397} {\bibfield  {journal} {\bibinfo  {journal} {J. Membr.
  Biol.}\ }\textbf {\bibinfo {volume} {164}},\ \bibinfo {pages} {103--114}
  (\bibinfo {year} {1998})}\BibitemShut {NoStop}%
\bibitem [{\citenamefont {Bleecker}\ \emph {et~al.}(2016)\citenamefont
  {Bleecker}, \citenamefont {Cox}, \citenamefont {Foster}, \citenamefont
  {Litz}, \citenamefont {Blosser}, \citenamefont {Castner},\ and\ \citenamefont
  {Keller}}]{Bleecker2016}%
  \BibitemOpen
  \bibfield  {author} {\bibinfo {author} {\bibfnamefont {J.~V.}\ \bibnamefont
  {Bleecker}}, \bibinfo {author} {\bibfnamefont {P.~A.}\ \bibnamefont {Cox}},
  \bibinfo {author} {\bibfnamefont {R.~N.}\ \bibnamefont {Foster}}, \bibinfo
  {author} {\bibfnamefont {J.~P.}\ \bibnamefont {Litz}}, \bibinfo {author}
  {\bibfnamefont {M.~C.}\ \bibnamefont {Blosser}}, \bibinfo {author}
  {\bibfnamefont {D.~G.}\ \bibnamefont {Castner}}, \ and\ \bibinfo {author}
  {\bibfnamefont {S.~L.}\ \bibnamefont {Keller}},\ }\bibfield  {title}
  {\enquote {\bibinfo {title} {Thickness mismatch of coexisting liquid phases
  in noncanonical lipid bilayers},}\ }\href@noop {} {\bibfield  {journal}
  {\bibinfo  {journal} {J. Phys. Chem. B}\ }\textbf {\bibinfo {volume} {120}},\
  \bibinfo {pages} {2761--2770} (\bibinfo {year} {2016})}\BibitemShut {NoStop}%
\bibitem [{\citenamefont {Dotsenko}(1995)}]{Dotsenko1995}%
  \BibitemOpen
  \bibfield  {author} {\bibinfo {author} {\bibfnamefont {V.~S.}\ \bibnamefont
  {Dotsenko}},\ }\bibfield  {title} {\enquote {\bibinfo {title} {Critical
  phenomena and quenched disorder},}\ }\href {\doibase
  10.1070/pu1995v038n05abeh000084} {\bibfield  {journal} {\bibinfo  {journal}
  {Phys.-Uspekhi}\ }\textbf {\bibinfo {volume} {38}},\ \bibinfo {pages}
  {457--497} (\bibinfo {year} {1995})}\BibitemShut {NoStop}%
\bibitem [{\citenamefont {Dotsenko}(2007)}]{Dotsenko2007}%
  \BibitemOpen
  \bibfield  {author} {\bibinfo {author} {\bibfnamefont {V.~S.}\ \bibnamefont
  {Dotsenko}},\ }\bibfield  {title} {\enquote {\bibinfo {title} {On the nature
  of the phase transition in the three-dimensional random field {Ising}
  model},}\ }\href {\doibase 10.1088/1742-5468/2007/09/p09005} {\bibfield
  {journal} {\bibinfo  {journal} {J. of Stat. Mech.: Theory Exp.}\ }\textbf
  {\bibinfo {volume} {2007}},\ \bibinfo {pages} {P09005} (\bibinfo {year}
  {2007})}\BibitemShut {NoStop}%
\bibitem [{\citenamefont {Grage}\ \emph {et~al.}(2016)\citenamefont {Grage},
  \citenamefont {Afonin}, \citenamefont {Kara}, \citenamefont {Buth},\ and\
  \citenamefont {Ulrich}}]{Grage2016}%
  \BibitemOpen
  \bibfield  {author} {\bibinfo {author} {\bibfnamefont {S.~L.}\ \bibnamefont
  {Grage}}, \bibinfo {author} {\bibfnamefont {S.}~\bibnamefont {Afonin}},
  \bibinfo {author} {\bibfnamefont {S.}~\bibnamefont {Kara}}, \bibinfo {author}
  {\bibfnamefont {G.}~\bibnamefont {Buth}}, \ and\ \bibinfo {author}
  {\bibfnamefont {A.~S.}\ \bibnamefont {Ulrich}},\ }\bibfield  {title}
  {\enquote {\bibinfo {title} {Membrane thinning and thickening induced by
  membrane-active amphipathic peptides},}\ }\href {\doibase
  10.3389/fcell.2016.00065} {\bibfield  {journal} {\bibinfo  {journal} {Front.
  Cell Dev. Biol.}\ }\textbf {\bibinfo {volume} {4}},\ \bibinfo {pages} {65}
  (\bibinfo {year} {2016})}\BibitemShut {NoStop}%
\bibitem [{Note1()}]{Note1}%
  \BibitemOpen
  \bibinfo {note} {For example, using estimates for coefficients from \cite
  {Bories2018}, the value of $\eta /\protect \sqrt {\kappa \gamma }$ does not
  exceed $0.5$.}\BibitemShut {Stop}%
\bibitem [{\citenamefont {Huang}(1986)}]{Huang1986}%
  \BibitemOpen
  \bibfield  {author} {\bibinfo {author} {\bibfnamefont {H.~W.}\ \bibnamefont
  {Huang}},\ }\bibfield  {title} {\enquote {\bibinfo {title} {Deformation free
  energy of bilayer membrane and its effect on gramicidin channel lifetime},}\
  }\href {\doibase https://doi.org/10.1016/S0006-3495(86)83550-0} {\bibfield
  {journal} {\bibinfo  {journal} {Biophys. J.}\ }\textbf {\bibinfo {volume}
  {50}},\ \bibinfo {pages} {1061--1070} (\bibinfo {year} {1986})}\BibitemShut
  {NoStop}%
\bibitem [{\citenamefont {Netz}(1997)}]{Netz1997}%
  \BibitemOpen
  \bibfield  {author} {\bibinfo {author} {\bibfnamefont {R.~R.}\ \bibnamefont
  {Netz}},\ }\bibfield  {title} {\enquote {\bibinfo {title} {Inclusions in
  fluctuating membranes: Exact results},}\ }\href {\doibase
  10.1051/jp1:1997205} {\bibfield  {journal} {\bibinfo  {journal} {J. Phys. I
  France}\ }\textbf {\bibinfo {volume} {7}},\ \bibinfo {pages} {833--852}
  (\bibinfo {year} {1997})}\BibitemShut {NoStop}%
\bibitem [{\citenamefont {Dommersnes}\ and\ \citenamefont
  {Fournier}(1999)}]{Dommersnes1999}%
  \BibitemOpen
  \bibfield  {author} {\bibinfo {author} {\bibfnamefont {P.}~\bibnamefont
  {Dommersnes}}\ and\ \bibinfo {author} {\bibfnamefont {J.-B.}\ \bibnamefont
  {Fournier}},\ }\bibfield  {title} {\enquote {\bibinfo {title} {{N}-body study
  of anisotropic membrane inclusions: Membrane mediated interactions and
  ordered aggregation},}\ }\href {\doibase 10.1007/s100510050968} {\bibfield
  {journal} {\bibinfo  {journal} {Eur. Phys. J. B}\ }\textbf {\bibinfo {volume}
  {12}},\ \bibinfo {pages} {9--12} (\bibinfo {year} {1999})}\BibitemShut
  {NoStop}%
\bibitem [{\citenamefont {Schmidt}\ \emph {et~al.}(2012)\citenamefont
  {Schmidt}, \citenamefont {Bihr}, \citenamefont {Seifert},\ and\ \citenamefont
  {Smith}}]{Schmidt2012}%
  \BibitemOpen
  \bibfield  {author} {\bibinfo {author} {\bibfnamefont {D.}~\bibnamefont
  {Schmidt}}, \bibinfo {author} {\bibfnamefont {T.}~\bibnamefont {Bihr}},
  \bibinfo {author} {\bibfnamefont {U.}~\bibnamefont {Seifert}}, \ and\
  \bibinfo {author} {\bibfnamefont {A.-S.}\ \bibnamefont {Smith}},\ }\bibfield
  {title} {\enquote {\bibinfo {title} {Coexistence of dilute and densely packed
  domains of ligand-receptor bonds in membrane adhesion},}\ }\href {\doibase
  10.1209/0295-5075/99/38003} {\bibfield  {journal} {\bibinfo  {journal} {EPL}\
  }\textbf {\bibinfo {volume} {99}},\ \bibinfo {pages} {38003} (\bibinfo {year}
  {2012})}\BibitemShut {NoStop}%
\bibitem [{\citenamefont {Jane\v{s}}\ \emph {et~al.}(2019)\citenamefont
  {Jane\v{s}}, \citenamefont {Stumpf}, \citenamefont {Schmidt}, \citenamefont
  {Seifert},\ and\ \citenamefont {Smith}}]{Janes2019}%
  \BibitemOpen
  \bibfield  {author} {\bibinfo {author} {\bibfnamefont {J.~A.}\ \bibnamefont
  {Jane\v{s}}}, \bibinfo {author} {\bibfnamefont {H.}~\bibnamefont {Stumpf}},
  \bibinfo {author} {\bibfnamefont {D.}~\bibnamefont {Schmidt}}, \bibinfo
  {author} {\bibfnamefont {U.}~\bibnamefont {Seifert}}, \ and\ \bibinfo
  {author} {\bibfnamefont {A.-S.}\ \bibnamefont {Smith}},\ }\bibfield  {title}
  {\enquote {\bibinfo {title} {Statistical mechanics of an elastically pinned
  membrane: Static profile and correlations},}\ }\href {\doibase
  https://doi.org/10.1016/j.bpj.2018.12.003} {\bibfield  {journal} {\bibinfo
  {journal} {Biophys. J.}\ }\textbf {\bibinfo {volume} {116}},\ \bibinfo
  {pages} {283--295} (\bibinfo {year} {2019})}\BibitemShut {NoStop}%
\bibitem [{\citenamefont {Fisher}\ and\ \citenamefont
  {Wiodm}(1969)}]{Fisher1969}%
  \BibitemOpen
  \bibfield  {author} {\bibinfo {author} {\bibfnamefont {M.~E.}\ \bibnamefont
  {Fisher}}\ and\ \bibinfo {author} {\bibfnamefont {B.}~\bibnamefont {Wiodm}},\
  }\bibfield  {title} {\enquote {\bibinfo {title} {Decay of correlations in
  linear systems},}\ }\href@noop {} {\bibfield  {journal} {\bibinfo  {journal}
  {J. Chem. Phys.}\ }\textbf {\bibinfo {volume} {50}},\ \bibinfo {pages}
  {3756--3772} (\bibinfo {year} {1969})}\BibitemShut {NoStop}%
\bibitem [{\citenamefont {Fisher}\ and\ \citenamefont
  {Widom}(2015)}]{Fisher2015}%
  \BibitemOpen
  \bibfield  {author} {\bibinfo {author} {\bibfnamefont {M.~E.}\ \bibnamefont
  {Fisher}}\ and\ \bibinfo {author} {\bibfnamefont {B.}~\bibnamefont {Widom}},\
  }\bibfield  {title} {\enquote {\bibinfo {title} {Publisher’s note: ``decay
  of correlations in linear systems''[{J. Chem. Phys.} 50, 3756 (1969)]},}\
  }\href@noop {} {\bibfield  {journal} {\bibinfo  {journal} {J. Chem. Phys.}\
  }\textbf {\bibinfo {volume} {143}},\ \bibinfo {pages} {209903} (\bibinfo
  {year} {2015})}\BibitemShut {NoStop}%
\bibitem [{\citenamefont {Evans}\ \emph {et~al.}(1994)\citenamefont {Evans},
  \citenamefont {Leote~de Carvalho}, \citenamefont {Henderson},\ and\
  \citenamefont {Hoyle}}]{Evans1994}%
  \BibitemOpen
  \bibfield  {author} {\bibinfo {author} {\bibfnamefont {R.}~\bibnamefont
  {Evans}}, \bibinfo {author} {\bibfnamefont {R.~J.~F.}\ \bibnamefont {Leote~de
  Carvalho}}, \bibinfo {author} {\bibfnamefont {J.~R.}\ \bibnamefont
  {Henderson}}, \ and\ \bibinfo {author} {\bibfnamefont {D.~C.}\ \bibnamefont
  {Hoyle}},\ }\bibfield  {title} {\enquote {\bibinfo {title} {Asymptotic decay
  of correlations in liquids and their mixtures},}\ }\href@noop {} {\bibfield
  {journal} {\bibinfo  {journal} {J. Chem. Phys.}\ }\textbf {\bibinfo {volume}
  {100}},\ \bibinfo {pages} {591--603} (\bibinfo {year} {1994})}\BibitemShut
  {NoStop}%
\bibitem [{\citenamefont {Nowakowski}\ \emph {et~al.}(2021)\citenamefont
  {Nowakowski}, \citenamefont {Stumpf}, \citenamefont {Smith},\ and\
  \citenamefont {Macio{\l}ek}}]{NSMS}%
  \BibitemOpen
  \bibfield  {author} {\bibinfo {author} {\bibfnamefont {P.}~\bibnamefont
  {Nowakowski}}, \bibinfo {author} {\bibfnamefont {B.~H.}\ \bibnamefont
  {Stumpf}}, \bibinfo {author} {\bibfnamefont {A.-S.}\ \bibnamefont {Smith}}, \
  and\ \bibinfo {author} {\bibfnamefont {A.}~\bibnamefont {Macio{\l}ek}},\
  }\href@noop {} {\enquote {\bibinfo {title} {Model for protein induced local
  demixing of binary lipid membranes},}\ } (\bibinfo
  {year} {2021}),\ \bibinfo {note} {in preparation}\BibitemShut {NoStop}%
\bibitem [{\citenamefont {Bihr}\ \emph {et~al.}(2015)\citenamefont {Bihr},
  \citenamefont {Seifert},\ and\ \citenamefont {Smith}}]{Bihr2015}%
  \BibitemOpen
  \bibfield  {author} {\bibinfo {author} {\bibfnamefont {T.}~\bibnamefont
  {Bihr}}, \bibinfo {author} {\bibfnamefont {U.}~\bibnamefont {Seifert}}, \
  and\ \bibinfo {author} {\bibfnamefont {A.-S.}\ \bibnamefont {Smith}},\
  }\bibfield  {title} {\enquote {\bibinfo {title} {Multiscale approaches to
  protein-mediated interactions between membranes—relating microscopic and
  macroscopic dynamics in radially growing adhesions},}\ }\href {\doibase
  10.1088/1367-2630/17/8/083016} {\bibfield  {journal} {\bibinfo  {journal}
  {New J. Phys.}\ }\textbf {\bibinfo {volume} {17}},\ \bibinfo {pages} {083016}
  (\bibinfo {year} {2015})}\BibitemShut {NoStop}%
\bibitem [{\citenamefont {Edison}\ \emph {et~al.}(2015)\citenamefont {Edison},
  \citenamefont {Tasios}, \citenamefont {Belli}, \citenamefont {Evans},
  \citenamefont {van Roij},\ and\ \citenamefont {Dijkstra}}]{Edison2015}%
  \BibitemOpen
  \bibfield  {author} {\bibinfo {author} {\bibfnamefont {J.~R.}\ \bibnamefont
  {Edison}}, \bibinfo {author} {\bibfnamefont {N.}~\bibnamefont {Tasios}},
  \bibinfo {author} {\bibfnamefont {S.}~\bibnamefont {Belli}}, \bibinfo
  {author} {\bibfnamefont {R.}~\bibnamefont {Evans}}, \bibinfo {author}
  {\bibfnamefont {R.}~\bibnamefont {van Roij}}, \ and\ \bibinfo {author}
  {\bibfnamefont {M.}~\bibnamefont {Dijkstra}},\ }\bibfield  {title} {\enquote
  {\bibinfo {title} {Critical casimir forces and colloidal phase transitions in
  a near-critical solvent: A simple model reveals a rich phase diagram},}\
  }\href {\doibase 10.1103/PhysRevLett.114.038301} {\bibfield  {journal}
  {\bibinfo  {journal} {Phys. Rev. Lett.}\ }\textbf {\bibinfo {volume} {114}},\
  \bibinfo {pages} {038301} (\bibinfo {year} {2015})}\BibitemShut {NoStop}%
\bibitem [{\citenamefont {Honerkamp-Smith}\ \emph {et~al.}(2008)\citenamefont
  {Honerkamp-Smith}, \citenamefont {Cicuta}, \citenamefont {Collins},
  \citenamefont {Veatch}, \citenamefont {{den Nijs}}, \citenamefont {Schick},\
  and\ \citenamefont {Keller}}]{Honerkamp-Smith2008}%
  \BibitemOpen
  \bibfield  {author} {\bibinfo {author} {\bibfnamefont {A.~R.}\ \bibnamefont
  {Honerkamp-Smith}}, \bibinfo {author} {\bibfnamefont {P.}~\bibnamefont
  {Cicuta}}, \bibinfo {author} {\bibfnamefont {M.~D.}\ \bibnamefont {Collins}},
  \bibinfo {author} {\bibfnamefont {S.~L.}\ \bibnamefont {Veatch}}, \bibinfo
  {author} {\bibfnamefont {M.}~\bibnamefont {{den Nijs}}}, \bibinfo {author}
  {\bibfnamefont {M.}~\bibnamefont {Schick}}, \ and\ \bibinfo {author}
  {\bibfnamefont {S.~L.}\ \bibnamefont {Keller}},\ }\bibfield  {title}
  {\enquote {\bibinfo {title} {Line tensions, correlation lengths, and critical
  exponents in lipid membranes near critical points},}\ }\href {\doibase
  https://doi.org/10.1529/biophysj.107.128421} {\bibfield  {journal} {\bibinfo
  {journal} {Biophys. J.}\ }\textbf {\bibinfo {volume} {95}},\ \bibinfo {pages}
  {236--246} (\bibinfo {year} {2008})}\BibitemShut {NoStop}%
\bibitem [{\citenamefont {Aranda-Espinoza}\ \emph {et~al.}(1996)\citenamefont
  {Aranda-Espinoza}, \citenamefont {Berman}, \citenamefont {Dan}, \citenamefont
  {Pincus},\ and\ \citenamefont {Safran}}]{ArandaEspiroza1996}%
  \BibitemOpen
  \bibfield  {author} {\bibinfo {author} {\bibfnamefont {H.}~\bibnamefont
  {Aranda-Espinoza}}, \bibinfo {author} {\bibfnamefont {A.}~\bibnamefont
  {Berman}}, \bibinfo {author} {\bibfnamefont {N.}~\bibnamefont {Dan}},
  \bibinfo {author} {\bibfnamefont {P.}~\bibnamefont {Pincus}}, \ and\ \bibinfo
  {author} {\bibfnamefont {S.}~\bibnamefont {Safran}},\ }\bibfield  {title}
  {\enquote {\bibinfo {title} {Interaction between inclusions embedded in
  membranes},}\ }\href {\doibase https://doi.org/10.1016/S0006-3495(96)79265-2}
  {\bibfield  {journal} {\bibinfo  {journal} {Biophys. J.}\ }\textbf {\bibinfo
  {volume} {71}},\ \bibinfo {pages} {648--656} (\bibinfo {year}
  {1996})}\BibitemShut {NoStop}%
\bibitem [{\citenamefont {Bories}\ \emph {et~al.}(2018)\citenamefont {Bories},
  \citenamefont {Constantin}, \citenamefont {Galatola},\ and\ \citenamefont
  {Fournier}}]{Bories2018}%
  \BibitemOpen
  \bibfield  {author} {\bibinfo {author} {\bibfnamefont {F.}~\bibnamefont
  {Bories}}, \bibinfo {author} {\bibfnamefont {D.}~\bibnamefont
  {Constantin}}, \bibinfo {author} {\bibfnamefont {P.}~\bibnamefont
  {Galatola}}, \ and\ \bibinfo {author} {\bibfnamefont {J.-B.}\ 
  \bibnamefont {Fournier}},\ }\bibfield  {title} {\enquote {\bibinfo {title}
  {Coupling between inclusions and membranes at the nanoscale},}\ }\href
  {\doibase 10.1103/PhysRevLett.120.128104} {\bibfield  {journal} {\bibinfo
  {journal} {Phys. Rev. Lett.}\ }\textbf {\bibinfo {volume} {120}},\ \bibinfo
  {pages} {128104} (\bibinfo {year} {2018})}\BibitemShut {NoStop}%
\end{thebibliography}

\begin{thebibliography}{3}%
\makeatletter
\providecommand \@ifxundefined [1]{%
 \@ifx{#1\undefined}
}%
\providecommand \@ifnum [1]{%
 \ifnum #1\expandafter \@firstoftwo
 \else \expandafter \@secondoftwo
 \fi
}%
\providecommand \@ifx [1]{%
 \ifx #1\expandafter \@firstoftwo
 \else \expandafter \@secondoftwo
 \fi
}%
\providecommand \natexlab [1]{#1}%
\providecommand \enquote  [1]{``#1''}%
\providecommand \bibnamefont  [1]{#1}%
\providecommand \bibfnamefont [1]{#1}%
\providecommand \citenamefont [1]{#1}%
\providecommand \href@noop [0]{\@secondoftwo}%
\providecommand \href [0]{\begingroup \@sanitize@url \@href}%
\providecommand \@href[1]{\@@startlink{#1}\@@href}%
\providecommand \@@href[1]{\endgroup#1\@@endlink}%
\providecommand \@sanitize@url [0]{\catcode `\\12\catcode `\$12\catcode
  `\&12\catcode `\#12\catcode `\^12\catcode `\_12\catcode `\%12\relax}%
\providecommand \@@startlink[1]{}%
\providecommand \@@endlink[0]{}%
\providecommand \url  [0]{\begingroup\@sanitize@url \@url }%
\providecommand \@url [1]{\endgroup\@href {#1}{\urlprefix }}%
\providecommand \urlprefix  [0]{URL }%
\providecommand \Eprint [0]{\href }%
\providecommand \doibase [0]{https://doi.org/}%
\providecommand \selectlanguage [0]{\@gobble}%
\providecommand \bibinfo  [0]{\@secondoftwo}%
\providecommand \bibfield  [0]{\@secondoftwo}%
\providecommand \translation [1]{[#1]}%
\providecommand \BibitemOpen [0]{}%
\providecommand \bibitemStop [0]{}%
\providecommand \bibitemNoStop [0]{.\EOS\space}%
\providecommand \EOS [0]{\spacefactor3000\relax}%
\providecommand \BibitemShut  [1]{\csname bibitem#1\endcsname}%
\let\auto@bib@innerbib\@empty
\bibitem [{\citenamefont {Honigmann}\ \emph {et~al.}(2014)\citenamefont
  {Honigmann}, \citenamefont {Sadeghi}, \citenamefont {Keller}, \citenamefont
  {Hell}, \citenamefont {Eggeling},\ and\ \citenamefont
  {Vink}}]{S10.7554/eLife.01671}%
  \BibitemOpen
  \bibfield  {author} {\bibinfo {author} {\bibfnamefont {A.}~\bibnamefont
  {Honigmann}}, \bibinfo {author} {\bibfnamefont {S.}~\bibnamefont {Sadeghi}},
  \bibinfo {author} {\bibfnamefont {J.}~\bibnamefont {Keller}}, \bibinfo
  {author} {\bibfnamefont {S.~W.}\ \bibnamefont {Hell}}, \bibinfo {author}
  {\bibfnamefont {C.}~\bibnamefont {Eggeling}},\ and\ \bibinfo {author}
  {\bibfnamefont {R.}~\bibnamefont {Vink}},\ }\href
  {https://doi.org/10.7554/eLife.01671} {\bibfield  {journal} {\bibinfo
  {journal} {eLife}\ }\textbf {\bibinfo {volume} {3}},\ \bibinfo {pages}
  {e01671} (\bibinfo {year} {2014})}\BibitemShut {NoStop}%
\bibitem [{\citenamefont {Bihr}\ \emph {et~al.}(2015)\citenamefont {Bihr},
  \citenamefont {Seifert},\ and\ \citenamefont {Smith}}]{SBihr2015}%
  \BibitemOpen
  \bibfield  {author} {\bibinfo {author} {\bibfnamefont {T.}~\bibnamefont
  {Bihr}}, \bibinfo {author} {\bibfnamefont {U.}~\bibnamefont {Seifert}},\ and\
  \bibinfo {author} {\bibfnamefont {A.-S.}\ \bibnamefont {Smith}},\ }\href
  {https://doi.org/10.1088/1367-2630/17/8/083016} {\bibfield  {journal}
  {\bibinfo  {journal} {New J. Phys.}\ }\textbf {\bibinfo {volume} {17}},\
  \bibinfo {pages} {083016} (\bibinfo {year} {2015})}\BibitemShut {NoStop}%
\bibitem [{\citenamefont {Nowakowski}\ \emph {et~al.}(2021)\citenamefont
  {Nowakowski}, \citenamefont {Stumpf}, \citenamefont {Smith},\ and\
  \citenamefont {Macio{\l}ek}}]{SNSMS}%
  \BibitemOpen
  \bibfield  {author} {\bibinfo {author} {\bibfnamefont {P.}~\bibnamefont
  {Nowakowski}}, \bibinfo {author} {\bibfnamefont {B.~H.}\ \bibnamefont
  {Stumpf}}, \bibinfo {author} {\bibfnamefont {A.-S.}\ \bibnamefont {Smith}},\
  and\ \bibinfo {author} {\bibfnamefont {A.}~\bibnamefont {Macio{\l}ek}}}
  (\bibinfo {year} {2021}),\ \bibinfo {note} {in preparation}\BibitemShut
  {NoStop}%
\end{thebibliography}
%

\pagebreak

\setcounter{page}{1}

\renewcommand{\theequation}{S\arabic{equation}}
\setcounter{equation}{0}

\renewcommand{\thefigure}{S\arabic{figure}}
\setcounter{figure}{0}

\onecolumngrid

\begin{center}
{\large \textbf{Supplemental Material for \\[0.08cm]
``Protein induced lipid demixing in homogeneous membranes''}}\\[0.4cm]
Bernd Henning Stumpf,$^\textrm{1}$ Piotr Nowakowski,$^{\textrm{2},\textrm{3}}$ Christian Eggeling,$^{\textrm{4},\textrm{5},\textrm{6}}$ Anna Macio\l{}ek,$^{\textrm{2},\textrm{7}}$ and Ana-Sun\v{c}ana Smith$^{\textrm{1},\textrm{8}}$\\[0.18cm]

\small$^\mathrm{1}$\textit{PULS Group, Institut f{\"u}r Theoretische Physik, IZNF,\\ Friedrich-Alexander-Universit{\"a}t Erlangen-N{\"u}rnberg, Cauerstra\ss{}e 3, 91058 Erlangen, Germany}\\
\small$^\mathrm{2}$\textit{Max-Planck-Institut f{\"u}r Intelligente Systeme Stuttgart, Heisenbergstr. 3, 70569 Stuttgart, Germany}\\
\small$^\mathrm{3}$\textit{Institut f{\"u}r Theoretische Physik IV, Universit{\"a}t Stuttgart, Pfaffenwaldring 57, 70569 Stuttgart}\\
\small$^\mathrm{4}$\textit{Institute of Applied Optics and Biophysics, Friedrich-Schiller-University Jena, 07743 Jena, Germany}\\
\small$^\mathrm{5}$\textit{Leibniz Institute of Photonic Technology e.V., 07745 Jena, Germany}\\
\small$^\mathrm{6}$\textit{MRC Human Immunology Unit and Wolfson Imaging Centre Oxford,\\ MRC Weatherall Institute of Molecular Medicine,\\ University of Oxford, Oxford OX3 9DS, United Kingdom}
\small$^\mathrm{7}$\textit{Institute of Physical Chemistry, Polish Academy of Sciences, Kasprzaka 44/52, PL-01-224 Warsaw, Poland}\\
\small$^\mathrm{8}$\textit{Group of Computational Life Sciences, Department of Physical Chemistry,\\
 Ru\dj{}er Bo\v{s}kovi\'c Institute, Bijeni\v{c}ka 54, 10000, Zagreb}
\end{center}

\section{Methods}
For details on the experiments see~\citep{S10.7554/eLife.01671}. In brief:
\subsection{Preparation of mica supported membranes}
Mica (Muscovite, Pelco, Ted Pella, Inc., 
Redding, CA) was cleaved into thin layers ($\sim\SI{10}{\micro\meter}$) and glued (optical UV 
adhesive No. 88, Norland Products Inc., Cranbury, NJ) onto clean glass cover slides. 
Immediately before spin-coating the lipid solution, the MICA on top of the glass 
was cleaved again to yield a thin ($\sim\SI{1}{\micro\meter}$) and clean layer. 
Next, $\SI{30}{\micro\litre}$ of $\SI{2}{\gram\per\litre}$ 
lipid solution in Methanol/Chloroform (1:1) were spin--coated ($\SI{2000}{\rpm}$, for $\SI{30}{\second}$) 
on top of the MICA. To remove residual solvent, the cover slide was put under 
vacuum for $\SI{20}{\minute}$. The supported lipid bilayer was hydrated with warm ($\SI{50}{\celsius}$) 
buffer ($\SI{150}{\milli\Molar}$ NaCl Tris pH 7.5) for $\SI{10}{\minute}$ and then rinsed several times to 
remove excess membranes until a single clean bilayer remained on the surface. 
All lipids were purchased from Avanti Polar Lipids, Inc., AL USA. The Ld phase 
was stained with far--red fluorescent DPPE--KK114 (Abberior GmbH). For imaging 
experiments, the concentration of fluorescent lipids was $\sim\SI{0.1}{\mpc}$; for FCS 
experiments $\sim\SI{0.01}{\mpc}$ was used.
\subsection{Actin binding to supported membranes}
Supported lipid bilayers were doped with biotinylated lipids 
\gls{dope-biotin}, \gls{dppe-biotin}, \gls{dspe-peg-biotin}, 
also purchased from Avanti) that were used to bind actin fibers to the membrane. 
The following procedure was performed at $\SI{37}{\celsius}$ to keep the membrane in the one 
phase region: The bilayer was incubated with $\SI{200}{\micro\litre}$ of $\SI{0.1}{\gram\per\litre}$ streptavidin 
for $\SI{10}{\minute}$ and then rinsed several times to remove unbound streptavidin. 
Next, the membrane was incubated with $\SI{200}{\micro\litre}$ of $\SI{1}{\micro\Molar}$ biotinylated phalloidin 
(Sigma--Aldrich, Steinheim, Germany) for $\SI{10}{\minute}$ and then rinsed several times to 
remove unbound phalloidin. Pre--polymerized actin fibers ($\SI{500}{\micro\litre}$ with $\SI{7}{\micro\gram\per\milli\litre}$ 
actin; Cytoskeleton Inc., Denver, USA) were then incubated with the membrane 
for $\SI{20}{\minute}$ and then rinsed several times to remove unbound actin. In case actin 
fibers were imaged, the actin was stained with green fluorescent phalloidin 
(Cytoskeleton Inc.). The membrane bound actin network was stable for at 
least $\SI{24}{\hour}$. The density of the actin network was controlled by the amount of 
biotinylated lipids in the membrane.
\subsection{Microscopy}
All fluorescence images were recorded on a custom--built confocal microscope, as 
highlighted in~\citep{S10.7554/eLife.01671}. Temperature was controlled using a water cooled Peltier heat and cooling stage which was mounted on the microscope (Warner Instruments, Hamden, CT, USA). The achievable temperature range with this configuration was between 7 and \SI{45}{\celsius}, with a precision of \SI{0.3}{\celsius}. The actual temperature directly over the membrane was measured by a small thermo-sensor (P605, Pt100, Dostmann electronic GmbH, Wertheim-Reicholzheim, Germany).

\begin{figure*}[htb]
\includegraphics[width=0.6\textwidth]{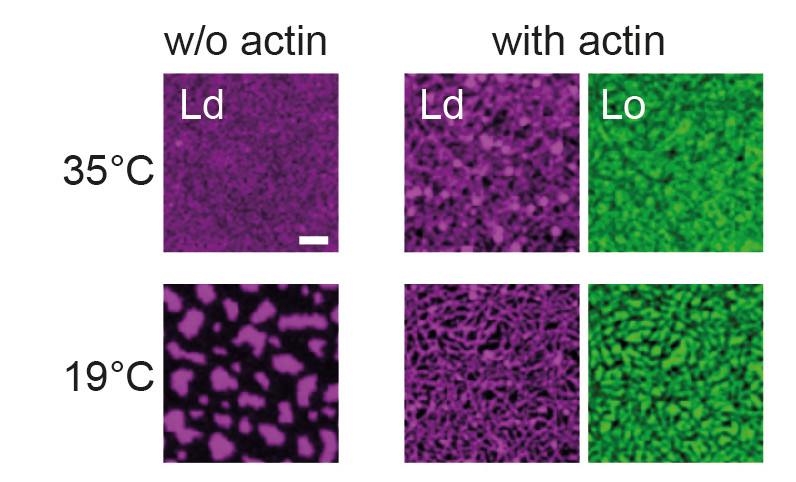}
\caption{\label{fig_control} 
Phase organization above and below the critical temperature with and without actin for the supported lipid bilayer with the same lipid mixture as in the main text, saturated lipids (DPPC), unsaturated lipids (DOPC) and cholesterol in a ratio of 35:35:30 mol\%, having a critical temperature $T_\mathrm{c} \approx \SI{28}{\celsius}$: confocal microscopy images of a fluorescent lipid preferring the Lo (DSPE--PEG--Chromeo488, green, Lo marker) and Ld phase (DPPE--KK114, red, Ld marker) without actin (left column) and with actin (two right columns) for $T = \SI{35}{\celsius}$ above $T_\mathrm{c}$ (upper row) and $T = \SI{19}{\celsius}$ below $T_\mathrm{c}$ (lower row). Without actin the supported lipid bilayer exhibits the typical phase separation behavior with coarsened domains below $T_\mathrm{c}$ and homogeneous membrane above $T_\mathrm{c}$, while after actin binding to DOPE--biotin the Ld domains form a meshwork--like structure and the Lo domains form an inverse pattern, similar to Fig. 1\textbf{A} in the main text, and the mosaic pattern persists above the phase transition temperature $T_\mathrm{c}$ of the membrane without actin. Scale bar $= 2 \si{\micro\meter}$. Adapted from Honigmann, A., Sadeghi, S., Keller, J., Hell, S.~W., Eggeling, C., and Vink, R.~(2014) \textit{A lipid bound actin meshwork organizes liquid phase separation in model membranes.}\ eLIFE 3:e01671.
}
\end{figure*}

\subsection{Control Images}

The phase separation along the actin fibres appears both above and below the critical temperature of the bare membrane.
In the main text, images for $T<T_\mathrm{c}$ are shown. However, the same effect appears at $T>T_\mathrm{c}$, see Fig.~\ref{fig_control}.

\section{Calculations}

In this section we give some details of our calculation.

\subsection{Dimensionless variables}

The Hamiltonian of the model of a membrane considered in the letter is
\begin{equation}\label{S:P:hamiltonianI}
\beta\mathcal{H}_0\left[\barh\left(\vecr\right),\barphi\left(\vecr\right)\right]=\int \mathrm{d}^2 r\Big[ \frac{\sigma}{2} \big( \nabla \barphi\left(\vecr\right)\!\big)^2+t\barphi^2\left(\vecr\right)+\frac{\gamma}{2} \big( \barh\left(\vecr\right)-\alpha \barphi\left(\vecr\right)\!\big)^2+\frac{\kappa}{2}\big( \nabla^2 \barh\left(\vecr\right)\!\big)^2\Big],
\end{equation}
where $\vecr$ is a position vector on 2D plane, $\barh\left(\vecr\right)$ is an order parameter describing the excess height of the membrane, $\barphi\left(\vecr\right)$ is an order parameter describing the composition of the membrane, the parameter $
\sigma$ measures the energy cost of inhomogeneous composition, $t$ is the reduced temperature (it is proportional to the difference between the temperature of the system $T$ and the critical temperature $T_\mathrm{c}$), $\alpha$ is the coupling between the order parameters, $\gamma$ measures the energy cost of violating the coupling relation $\barh\left(\vecr\right)=\alpha\barphi\left(\vecr\right)$, and $\kappa$ is the bending stiffness of the membrane.

The position vector $\vecr$ and the order parameter $\barh$ are measured in the units of length, which we denote by $\mathrm{L}$. The composition order parameter $\barphi$ is defined as a difference between local concentration of saturated lipids and their concentration at the critical demixing point, and, depending on the type of concentration that is used, it can be measured in $\mathrm{mol/m^2}$, $1/\mathrm{m}^2$, $\mathrm{kg/m^2}$, \textit{etc}. We denote this unit by $\mathrm{C}$. (We note that, close to $T_\mathrm{c}$ all possible definitions $\barphi$ differ only by a dimensional factor, which rescales the parameters $\sigma$, $t$ and $\alpha$. This way the properties of the system do not depend on the exact definition of $\barphi$.)

The analysis of Eq.~\eqref{S:P:hamiltonianI} shows that the parameter $\sigma$ is measured in units $\mathrm{C}^{-2}$, $t$ in $\mathrm{C}^{-2}\mathrm{L}^{-2}$, $\gamma$ in $\mathrm{L}^{-4}$, $\alpha$ in $\mathrm{L}/\mathrm{C}$, and $\kappa$ is dimensionless. We take $\zeta=\left(\kappa/\gamma\right)^{1/4}$ as a unit of length and $\sigma^{-1/2}$ as a unit of concentration, and we rescale all quantities present in our model to make them dimensionless. The dimensionless position is $\vecrho=\vecr/\zeta$ and the dimensionless order parameters are
\begin{equation}
    h\left(\vecrho\right)=\barh\left(\vecrho\zeta\right)/\zeta,\qquad \phi\left(\vecrho\right)=\barphi\left(\vecrho\zeta\right) \sigma^{1/2}.
\end{equation}
We also rescale the parameters $t$ and $\alpha$:
\begin{equation}
    \tau=t \zeta^2/\sigma, \qquad \mu=\alpha/\left(\zeta \sigma^{1/2}\right),
\end{equation}
which we call dimensionless (rescaled) temperature and dimensionless coupling, respectively.

Using the dimensionless position, fields, and parameters, the Hamiltonian in Eq.~\eqref{S:P:hamiltonianI} can be simplified
\begin{equation}
\beta\mathcal{H}_0\left[h\left(\vecrho\right),\phi\left(\vecrho\right)\right]=\int \mathrm{d}^2 \rho\Big[ \frac{1}{2} \big( \nabla \phi\left(\vecrho\right)\!\big)^2+\tau\phi^2\left(\vecrho\right)+\frac{\kappa}{2} \big( h\left(\vecrho\right)-\mu \phi\left(\vecrho\right)\!\big)^2+\frac{\kappa}{2}\big( \nabla^2 
h\left(\vecrho\right)\!\big)^2\Big].
\label{eq:Hz}
\end{equation}
This way, the bulk system is described by only three dimensionless parameters: $\kappa$, $\mu$ and $\tau$.

The interaction of the membrane with the anchoring proteins is given by
\begin{equation}\label{eq:Hintdim}
    \beta \mathcal{H}_\text{int}=\frac{\lambda}{2}\sum_{i=1}^N \big(\barh\left(\vecr_i\right)-\barh_0\big)^2,
\end{equation}
where $\vecr_i$ for $i=1,\ldots,N$ are the anchoring points and $\barh_0$ is the excess thickness of the membrane in the anchoring points. Changing the variables in the Hamiltonian~\eqref{eq:Hintdim} to make them dimensionless, we get
\begin{equation}\label{eq:Hint}
    \beta \mathcal{H}_\text{int}=\frac{\eta}{2}\sum_{i=1}^N \big(h\left(\vecrho_i\right)-h_0\big)^2,
\end{equation}
where $\vecrho_i=\vecr_i/\zeta$, $h_0=\barh_0/\zeta$, and $\eta=\lambda\zeta^2$. We note, that in our calculation we take the limit $\lambda\to\infty$, which is equivalent to $\eta\to\infty$.

\subsection{Correlation functions and profiles of the order parameters}

The calculation of the order parameter profiles follows the approach in~\citep{SBihr2015} with the additional composition order parameter $\phi\left(\vecrho\right)$, and it will be shown in more detail in~\citep{SNSMS}. Here, we only sketch the main steps of the calculation.



In the canonical ensemble the density of the probability distribution of having $h\left(\vecrho_\mathrm{a}\right)=\lambda_\mathrm{a}$, $h\left(\vecrho_\mathrm{b}\right)=\lambda_\mathrm{b}$ and $\phi\left(\vecrho_\mathrm{a}\right)=\vartheta_\mathrm{a}$ for two arbitrary vectors $\vecrho_\mathrm{a}$ and $\vecrho_\mathrm{b}$ in 2D plane is given by
\begin{multline}\label{eq:p}
    p(\kappa, \mu, \tau; h_0, \{\vecrho_i\}; \vecrho_\mathrm{a}, \lambda_\mathrm{a}, \theta_\mathrm{a}, \vecrho_\mathrm{b}, \lambda_\mathrm{b}) = \\
    \lim_{A \rightarrow \infty} C \int \mathcal{D}h\left(\vecrho\right) \mathcal{D}\phi\left(\vecrho\right) \delta \left[h\left(\vecrho_\mathrm{a}\right) - \lambda_\mathrm{a}\right]\delta \left[\phi\left(\vecrho_\mathrm{a}\right) - \theta_\mathrm{a}\right]\delta \left[h\left(\vecrho_\mathrm{b}\right) - \lambda_\mathrm{b}\right] \exp \left[ - \beta \mathcal{H}\left(h\left(\vecrho\right), \phi\left(\vecrho\right)\right) \right ],
\end{multline}
where $\vecrho_i$ for $i=1,\ldots, N$ denotes the points where the membrane is anchored, $h_0$ is the excess thickness enforced in the points of the anchors, $A$ is an area of the membrane, and $\mathcal{H}=\mathcal{H}_0+\mathcal{H}_\text{int}$ is given by Eqs.~\eqref{eq:Hz} and \eqref{eq:Hint}. In Eq.~\eqref{eq:p}, the function $\delta\left(x\right)$ is a Dirac delta, the symbol $\int \mathcal{D}h\left(\vecrho\right) \mathcal{D}\phi\left(\vecrho\right)$ denotes a path integral over all configurations of the order parameters and the constant $C$ is determined from the normalization condition
\begin{equation}\label{eq:normalization}
    \int_{-\infty}^{\infty}\dd \lambda_\mathrm{a} \int_{-\infty}^\infty \dd \lambda_\mathrm{b} \int_{-\infty}^\infty \dd \theta_\mathrm{a}\, p\left(\kappa, \mu, \tau; h_0, \{\vecrho_i\}; \vecrho_\mathrm{a}, \lambda_\mathrm{a}, \theta_\mathrm{a}, \vecrho_\mathrm{b}, \lambda_\mathrm{b}\right)=1.
\end{equation}

Following \cite{SBihr2015}, we calculate $p$ given by Eq.~\eqref{eq:p} in the following steps: First, we replace all Dirac delta functions using the relation
\begin{equation}
    \delta\left(x\right)=\frac{1}{2\pi}\int_{-\infty}^\infty \dd \psi \exp\left(\iu \psi x\right), 
\end{equation}
and the anchoring terms from $\mathcal{H}_\text{int}$ using a Hubbard--Stratonovich transformation. Second, we do the Fourier transformation for both order parameters; after this operation the formula for $p$ is a product of Gaussian integrals. In the third step, we do all these integrals, separately for every wavewector. Finally, we take the limit $A\to\infty$ and determine the constant $C$ from the normalization condition \eqref{eq:normalization}.

Using $p$, we calculate the correlation functions for a membrane without anchors
\begin{subequations}\label{eq:results}
\begin{align}
\corrfunc_{hh}\left(\rho; \kappa, \mu, \tau\right)&=\left<h\left(\vecrho_\mathrm{a}\right)h\left(\vecrho_\mathrm{b}\right)\right>=\int_{-\infty}^{\infty}\dd \lambda_\mathrm{a} \int_{-\infty}^\infty \dd \lambda_\mathrm{b} \int_{-\infty}^\infty \dd \theta_\mathrm{a}\,\lambda_\mathrm{a}\lambda_\mathrm{b} p(\kappa, \mu, \tau; \vecrho_\mathrm{a}, \lambda_\mathrm{a}, \theta_\mathrm{a}, \vecrho_\mathrm{b}, \lambda_\mathrm{b}),\\
\corrfunc_{h\phi}\left(\rho; \kappa, \mu, \tau\right)&=\left<\phi\left(\vecrho_\mathrm{a}\right)h\left(\vecrho_\mathrm{b}\right)\right>=\int_{-\infty}^{\infty}\dd \lambda_\mathrm{a} \int_{-\infty}^\infty \dd \lambda_\mathrm{b} \int_{-\infty}^\infty \dd \theta_\mathrm{a}\,\theta_\mathrm{a}\lambda_\mathrm{b} p(\kappa, \mu, \tau; \vecrho_\mathrm{a}, \lambda_\mathrm{a}, \theta_\mathrm{a}, \vecrho_\mathrm{b}, \lambda_\mathrm{b}),
\end{align}
where, due to the rotational symmetry of the system, the correlation functions depend on the distance $\rho=\left|\vecrho_\mathrm{b}-\vecrho_\mathrm{a}\right|$ rather than on the exact position of two points, and we have used the fact that for a membrane without anchors $\left<h\left(\vecrho_\mathrm{a}\right)\right>=\left<h\left(\vecrho_\mathrm{b}\right)\right>=\left<\phi\left(\vecrho_\mathrm{a}\right)\right>=0$. The average values of the order parameters are given by
\begin{align}
\left< h(\vecrho_\mathrm{a})\right>&= \int_{-\infty}^{\infty}\dd \lambda_\mathrm{a} \int_{-\infty}^\infty \dd \lambda_\mathrm{b} \int_{-\infty}^\infty \dd \theta_\mathrm{a}\, \lambda_\mathrm{a} p(\kappa, \mu, \tau; h_0, \{\vecrho_i\}; \vecrho_\mathrm{a}, \lambda_\mathrm{a}, \theta_\mathrm{a}, \vecrho_\mathrm{b}, \lambda_\mathrm{b}),\\
\left< \phi(\vecrho_\mathrm{a})\right>&= \int_{-\infty}^{\infty}\dd \lambda_\mathrm{a} \int_{-\infty}^\infty \dd \lambda_\mathrm{b} \int_{-\infty}^\infty \dd \theta_\mathrm{a}\, \theta_\mathrm{a} p(\kappa, \mu, \tau; h_0, \{\vecrho_i\}; \vecrho_\mathrm{a}, \lambda_\mathrm{a}, \theta_\mathrm{a}, \vecrho_\mathrm{b}, \lambda_\mathrm{b}).
\end{align}
\end{subequations}
Calculating the formulae in Eq.~\eqref{eq:results} gives the equations for the correlation functions and average order parameters in the main text.

\end{document}